\documentclass{aa}
\usepackage{graphicx}
\usepackage[varg]{txfonts}
\usepackage{lscape}

\begin{document}
\titlerunning{Discovery of a new branch of the Taurid meteoroid stream}
\title{Discovery of a new branch of the Taurid meteoroid stream as a real source of potentially hazardous bodies}

 \author{
P. Spurn\'y\inst{1}  \and J. Borovi\v{c}ka\inst{1}
\and H. Mucke\inst{2} \and J. Svore\v{n}\inst{3}
}

   \institute{Astronomical Institute of the Czech Academy of Sciences, CZ-25165 Ond\v{r}ejov, Czech Republic \\
              \email{pavel.spurny@asu.cas.cz}
   \and
  Astronomisches B\" uro, 1230 Wien, Austria
  \and
  Astronomical Institute of the Slovak Academy of Sciences, SK-05960 Tatransk\' a Lomnica, Slovak Republic
 }

\date{Received March 14, 2017; accepted May 1, 2017}

\abstract{Taurid meteor shower produces prolonged but usually low activity every October and November. 
In some years, however, the activity
is significantly enhanced. Previous studies based on long-term activity statistics concluded that the enhancement is caused
by a swarm of meteoroids locked in 7:2 resonance with Jupiter. Here
we present precise data on 144 Taurid fireballs observed by new digital cameras of the European Fireball Network 
in the enhanced activity year 2015. 
Orbits of 113 fireballs show common characteristics and form together a well defined orbital structure, which we call
new branch and which was evidently responsible for the enhanced activity. This new branch is part of Southern Taurids and 
was encountered by the Earth between October 25 and November 17. We found that this branch is characterized by longitudes of perihelia 
lying between 155.9 -- 160$\degr$ and latitudes of perihelia between
4.2 -- 5.7$\degr$. Semimajor axes are between 2.23 -- 2.28 AU and indeed overlap with the 7:2 resonance. 
Eccentricities are in wide range 0.80 -- 0.90.  The most eccentric orbits with lowest perihelion distances 
were encountered at the beginning of the activity period. The orbits form a concentric ring in the inner solar system.
The masses of the observed meteoroids were in a wide range from 0.1 gram to more than 1000 kg.  We found that
all meteoroids larger than 300 grams were very fragile (type IIIB), while those smaller than 30 grams were much more compact
(mostly of type II and some of them even type I). 
Based on orbital characteristics, we argue that asteroids 2015 TX24 and 2005 UR, both of 
diameters 200 -- 300 meters, are direct members of the new branch. 
It is therefore very likely that the new branch contains also numerous still not discovered objects of decameter or even larger size. 
Since asteroids of sizes of tens to hundreds meters pose a treat to the ground even if they are intrinsically weak, impact hazard
increases significantly when the Earth encounters the Taurid new branch every few years. Further studies leading to better description of this real source of potentially hazardous objects, which can be large enough to cause significant regional or even continental damage on the Earth, are therefore extremely important.
}


\maketitle
\section{Introduction}

The Taurid meteoroid stream is one of the most studied meteoroid streams. This stream produces at least four meteor showers
on Earth: the Northern and Southern Taurids, both active from end of September until December; the Daytime $\zeta$-Perseids,
active from end of May to the beginning of July; and the Daytime $\beta$-Taurids, active in June and the first half of July \citep{Jennbook}.
Other showers may be also related to the Taurid stream, namely the Piscids in September, $\chi$-Orionids in December, and Daytime May Arietids
in May \citep{Jennbook}. Since the work of \citet{Whipple}, the short period comet 2P/Encke has been considered the most
probable parent body of the Taurid stream. It was, nevertheless,  proposed that 2P/Encke is just a fragment of a much larger comet, which was disrupted
$10^3$--$10^4$ years ago and formed the whole Taurid complex including a number of asteroids \citep{Clube, Napier}.
As more and more asteroids were being discovered over time, the number of asteroids proposed by various authors
as members of the Taurid complex increased \citep[e.g.,][]{Asher93, Babadz01, Porub06, Babadz08, Olech}.
The problem is that the Taurid stream is very extended and the low-inclination short-period Taurid orbits are very common
for near-Earth asteroids. Many proposed associations can be therefore just random coincidences. Indeed, the spectra
of six large asteroids proposed as members of the Taurid complex showed that five of them are inconsistent with a cometary
origin \citep{Popescu}.

\begin{figure*}
\centering
\includegraphics[width=0.9\hsize]{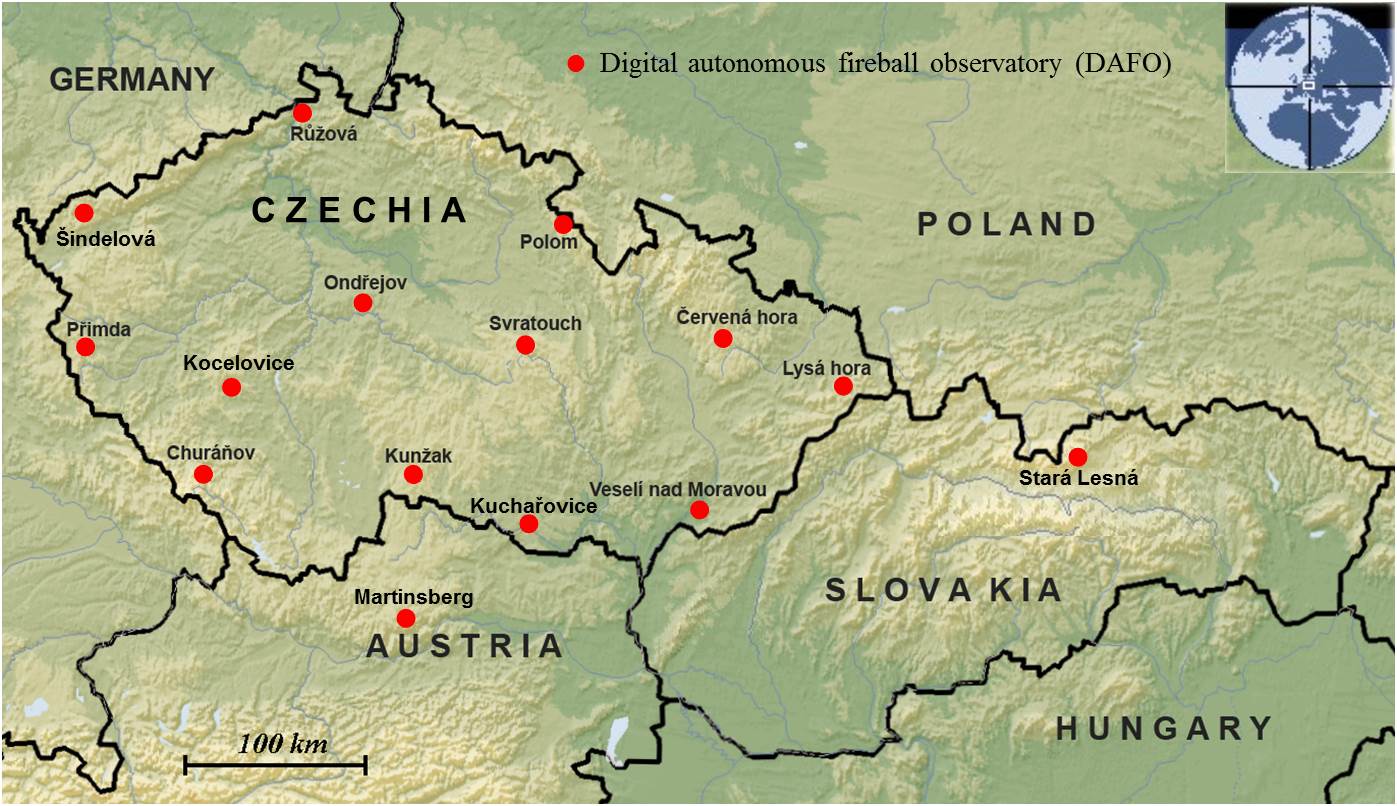}
\caption{Stations of the fireball network located in Czech Republic, Slovakia and Austria where DAFO are placed (status November 2015)}
\label{network}
\end{figure*}

The activity of Taurids is prolonged but usually of low level. In some years, however, the activity is enhanced, especially
in terms of large numbers of bright meteors (fireballs). \citet{QJRAS} proposed that there is a resonant swarm of meteoroids
trapped in the 7:2 resonance with Jupiter. The expected extent of the swarm was $\pm$ 30 -- $40\degr$ in mean anomaly. \citet{Asher98}
showed that enhanced Taurid activity indeed occurred in the years when the center of the swarm was less than $40\degr$ in mean
anomaly from the Earth at the beginning of November (the date of Taurid maximum). \citet{QJRAS} predicted that future
encounters would occur in 1995, 1998, 2005, and 2008. In 1995, enhanced Taurid activity was observed by the European Fireball
Network (EN)  when the rate of registered Taurid fireballs was noticeably higher than is usual for the EN at that time of year \citep{Spurny96}.
Apart from regular Southern and Northern Taurids, five fireballs observed during the last week of October 1995 had
distinct but very similar orbits. The radiants lay near the regular southern Taurid radiant, but the initial velocities
were larger ($V_g = 33.1 \pm 0.3$ km s$^{-1}$). As a result these orbits had significantly larger semimajor axes
($a = 2.52 \pm 0.08$ AU), eccentricities ($e = 0.905 \pm 0.004$) and inclinations ($i = 6.2^\circ \pm 0.4^\circ$),
and smaller perihelion distances ($q = 0.241 \pm 0.009$ AU) than the regular Southern Taurid orbit.
The existence of this well-defined cluster of similar Taurid meteoroids very probably means that the enhanced activity in 1995
was caused by a new relatively compact subsystem of the Taurid complex close to the Southern Taurids.
As mentioned in \citet{Jennbook}, this observation identifies, for the first time, a meteor outburst associated with the Taurid shower,
and by implication the Earth crossing a relatively young dust trail. The 1995 enhanced activity was also confirmed
by visual observations \citep{Dubietis}, both in terms of increased overall activity and increased percentage of fireballs
in the period from October 23 to November 15 \citep{McBeath}. Similarly \citet{Johannink}, also from visual observations,
confirmed increased Taurid activity in 1998 and 2005 and suggested that the Southern Taurids are responsible
for the higher activity in resonance years.

\citet{Shiba} analyzed Taurid video observations from 2007 to 2015, including swarm encounter years 2008, 2012, and 2015
(see the webpage of D. Asher\footnote{http://star.arm.ac.uk/~dja/taurid/swarmyears.html} for swarm encounter predictions). He
confirmed that the enhanced activity is exclusively due to Southern Taurids. Shiba also studied the dependency of orbital elements
on time and found that not only the mean orbital period of swarm meteoroids but also that of Northern Taurids correspond to the 7:2 resonance with Jupiter. The eccentricity was found to decrease and perihelion distance to increase with time.

The work of Shiba work is statistical in nature, involving thousands of meteors but with large individual uncertainties.
Here we present precise data on 144 Taurid fireballs observed by new digital cameras of the European Fireball Network in 2015.
The description of the observational system, examples of the data, and demonstration of their precision are given
in Sect.~\ref{observations} and \ref{methods}. In Sect.~\ref{data} we show that the enhanced activity in 2015 was caused
by a well-defined branch of Taurid meteoroids.
We concentrate our study on orbital elements and only briefly discuss the physical properties of the meteoroids.
 In Sect.~\ref{asteroids} we show that several known asteroids also belong to the branch, which caused the 2015 activity.
The implications of our work are discussed in Sect.~\ref{discussion}.

\section{Observational techniques and data acquisition}
\label{observations}

The data reported here were obtained by the European Fireball Network (EN). The core of the network, located in the
Czech Republic, has been modernized  several times \citep{spu07}. But the last significant improvement has been
realized during the last three years when a completely new instrument, the high-resolution digital autonomous fireball
observatory (DAFO), was developed and gradually installed on the stations of the fireball network between
November 2013 and September 2015. These new all-sky digital cameras are working alongside the older analog
(using photographic films) autonomous all-sky cameras (AFO) on the majority of Czech stations but this older system based on
AFOs is gradually being decommissioned. At the end of 2015 the DAFOs were installed on 13 stations  around the
Czech Republic (\v Sindelov\' a and Kocelovice stations are completely new and were built in mid-2015).
Apart from the Czech territory, two DAFOs were installed on the already working stations in Slovakia and Austria, respectively.
The first DAFO\ was installed at the observatory of the Slovak Academy of Sciences in Tatransk\' a Lomnica, where one AFO also remains
in full operation, and the second, installed at the Waldviertel Observatory in Martinsberg (Austria),  substituted
the previous AFO system in September 2015. This core of the EN as schematically shown in Figure~\ref{network} also cooperates
with other parts and systems located in neighboring European countries but data used in this study are solely acquired
by the stations based on the DAFO (vast majority of used records) and AFO cameras as described above.

\begin{figure*}
\centering
\includegraphics[width=0.9\hsize]{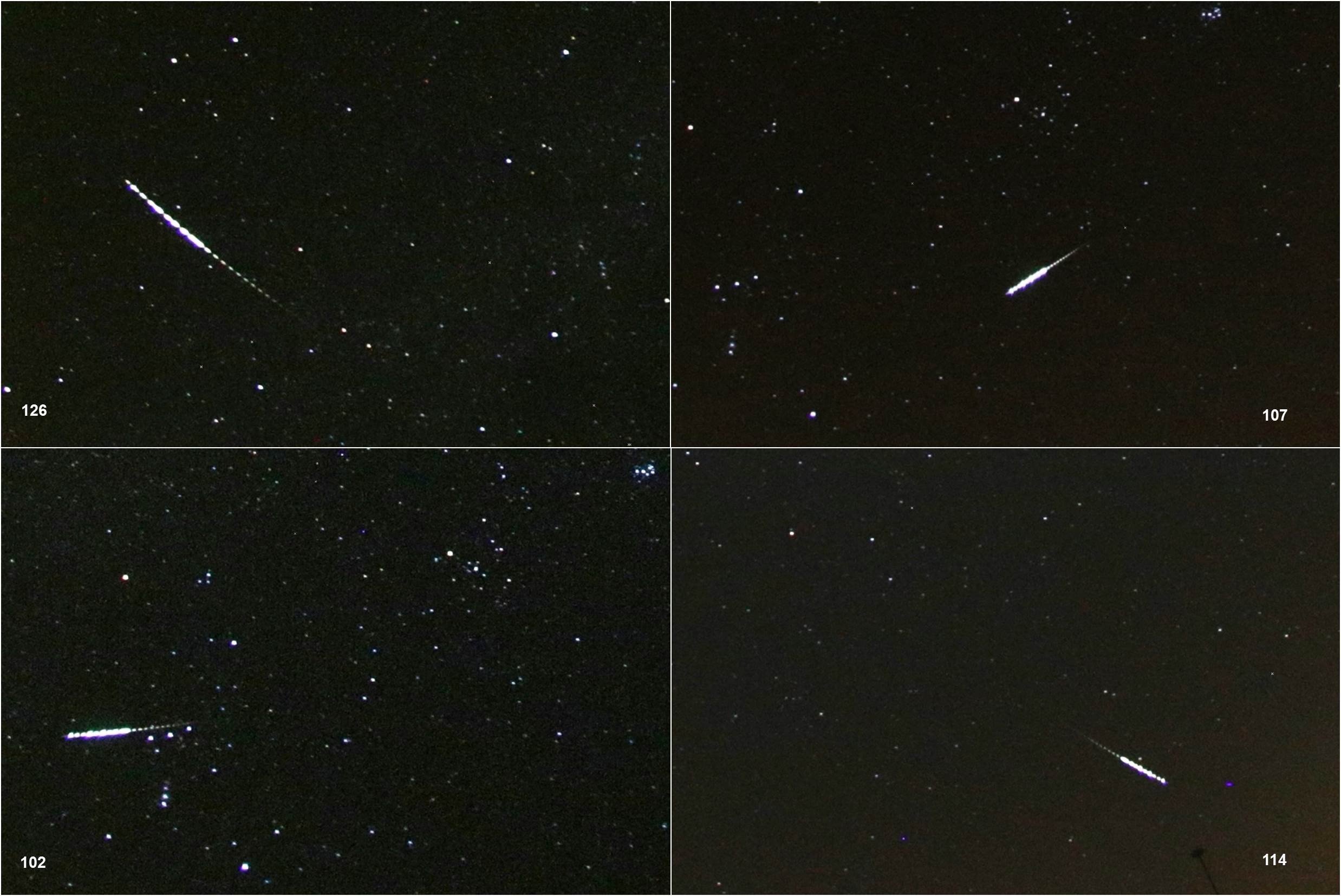}
\caption{Detailed views of the EN051115\_231201 Taurid fireball recorded by DAFOs at the stations 126 Martinsberg,
102 Kun\v zak, 107 Kucha\v rovice, and 114 \v Cerven\' a hora. All-sky images from these stations were used for the analysis.}
\label{firexample}
\end{figure*}

The imaging part of the DAFO system is comprised of a full frame Canon 6D digital camera and a Sigma fish-eye lens (8 mm f/3.5)
equipped with an electronic LCD shutter for speed determination.
In standard regime 16 interruptions and 35 s long exposure are used. To avoid possible loss of data during reading time
of the CMOS sensor, we use two identical imaging sets, which work in alternation mode with 5 second overlap.
The older AFOs analog imaging part is comprised of a Zeiss Distagon fish-eye lens (30 mm f/3.5). Large format panchromatic
sheet films (9 x 12 cm, Ilford FP4) are used. The diameter of the sky on the image is 8 cm and usually one exposure is taken per night.
Mechanical shutter with 15 interruptions per second is used. The sensitivity limit is $-$4 magnitude for AFO (about 2$-$3 mag
lower around the full Moon period) and $-$2 magnitude for DAFO (with lower dependence on lunar phase).
Apart from the imaging part, each DAFO and AFO is equipped with an all-sky radiometer with time resolution
of 5000 samples per second and with similar sensitivity limit (in the moonless nights) like the imaging system
but with much higher dynamic range. These radiometers serve several purposes, such as the real-time detection of fireballs,
their exact absolute timing (system time is continuously corrected by the PPS pulse of the GPS), recording
of detailed light curve profiles, and for precise photometry, especially for brighter events when digital images
become saturated as shown in one example later.

The data presented in this study were obtained almost completely by the new digital autonomous system (DAFO).
Thanks to their higher sensitivity, fireball observations from DAFO contain more information especially in the beginning and terminal parts of the
luminous trajectory in comparison with AFO. Another  important advantage of DAFO is the ability to work  during periods
when it is not completely dark (twilight periods) and not completely clear (partly cloudy sky) as well. The data from the new digital system
allow us to reliably determine all basic parameters of sufficiently bright fireballs up to the distance of 300 km from the stations
(for special cases even up to 600 km). It means that with the current number and displacement of stations (see Figure~\ref{network})
we effectively cover territory of roughly 3 million square kilometers at least, i.e., a large part of Central Europe. All the advantages
mentioned above significantly increased the efficiency of our observations; and in direct comparison with the efficiency
of the previous analog AFO system the number of recorded fireballs increased at least three times. When we combine this
increased efficiency with improved analysis techniques, which we developed and gradually improved
especially in the last several years, we obtain results than were not reached by any previous
observing system used within the EN.

\begin{figure}
\centering
\includegraphics[width=0.9\hsize]{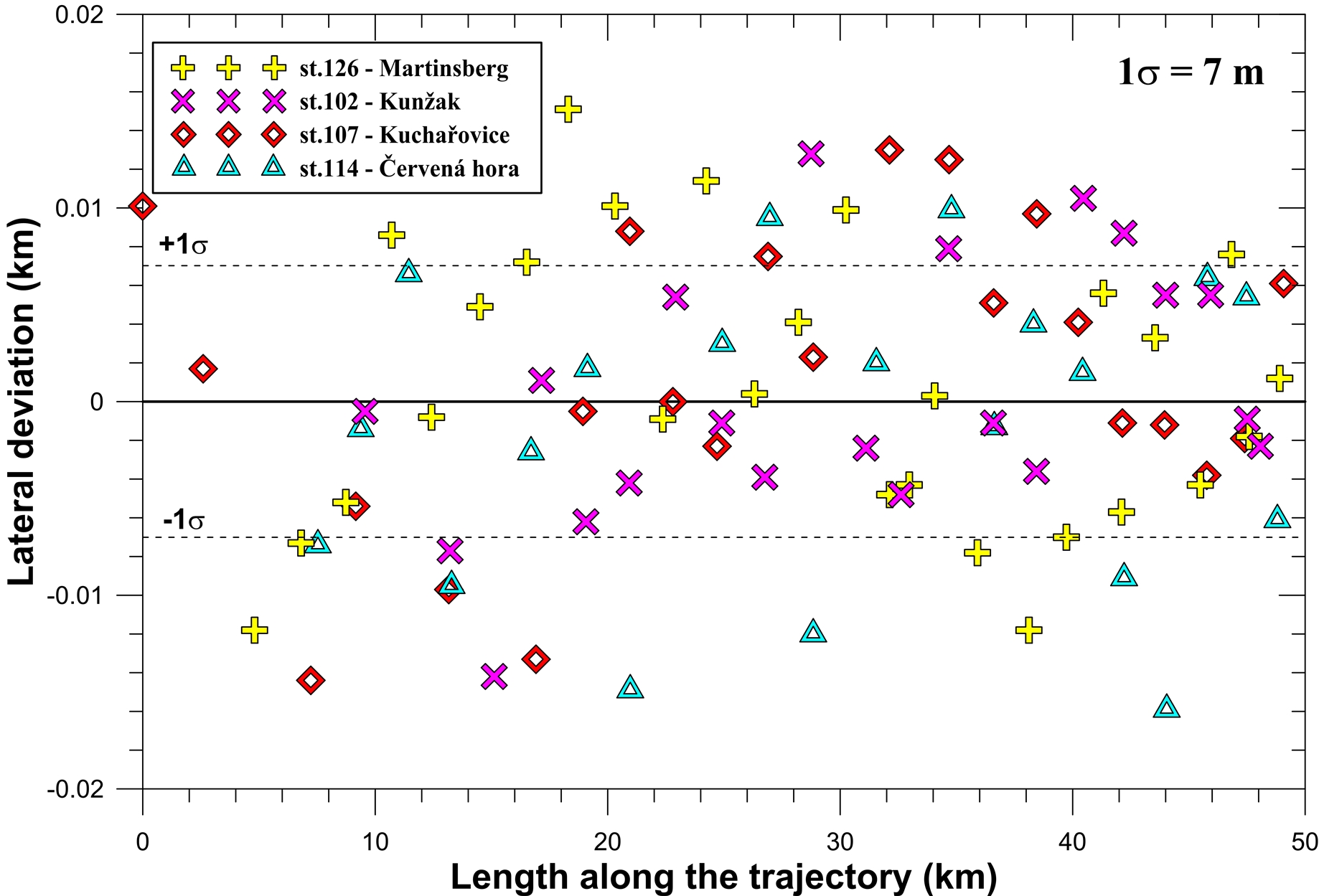}
\caption{Lateral deviations of all measured points on the fireball luminous path from the available records.
The Y-axis scale is highly enlarged and one standard deviation for any point on the fireball trajectory is only 7 m.}
\label{trajectory}
\end{figure}

\section{Data reduction}
\label{methods}

As described above, our fully automated instruments DAFO and AFO provide us with two kinds of data: all-sky photographic
records and high-resolution radiometric light curves. For the complete analysis of every fireball that was recorded from
at least two stations (the vast majority of the presented fireballs were recorded from more than two stations) we use our own procedures,
methods and analysis software. All-sky images taken in the raw format are measured by the FishScan application, which allows semiautomatic measurement of positions, speed, and photometry. Usually the photometry from digital images  is reliable
up to $-8$ apparent magnitude, brighter events start to be saturated after reaching this brightness. However, thanks to the high
dynamic range of radiometers, which are incorporated in each DAFO and AFO, we are able to obtain precise photometry also
for much brighter fireballs, even for superbolides as will be shown later.  We can calibrate radiometric records using not
saturated parts of the light curve obtained from photographic records. Our whole procedure is demonstrated on the example below.

\begin{figure}
\centering
\includegraphics[width=0.9\hsize]{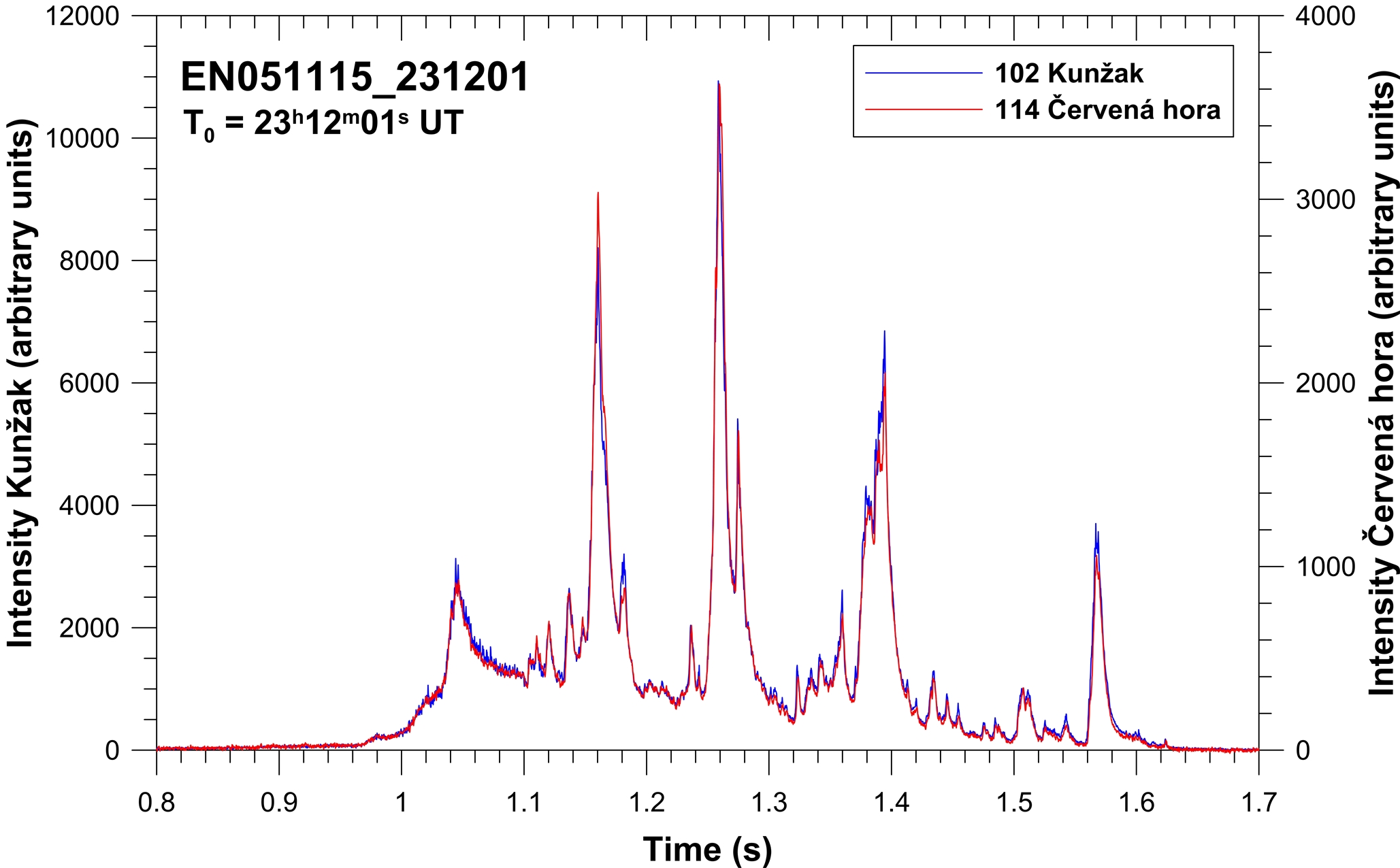}
\caption{Radiometric light curves of the EN051115\_231201 fireball taken by fast photometers
(5000 samples/s) at Kun\v zak (blue) and \v Cerven\' a hora (red) stations. These apparent (not corrected for distance)
light curves taken from places185 km apart demonstrate perfect compliance of both records; small differences in heights
of individual peaks are caused by different distances to the fireball.}
\label{twolcint}
\end{figure}

\subsection {EN051115\_231201 Taurid fireball: Example of data analysis}
\label{precision}

The Taurid fireball of November 5, 2015, 23:12:01 UT, was recorded photographically and photoelectrically at seven stations
in our network. For a complete analysis of this fireball, we chose records taken from four stations that were close to
its atmospheric trajectory and were sufficient for reliable  determination of all parameters describing atmospheric trajectory,
dynamics, photometry, and heliocentric orbit of this fireball. A selection of all-sky images of the fireball taken
at individual stations are shown in Figure~\ref{firexample}.

\begin{figure}
\centering
\includegraphics[width=0.9\hsize]{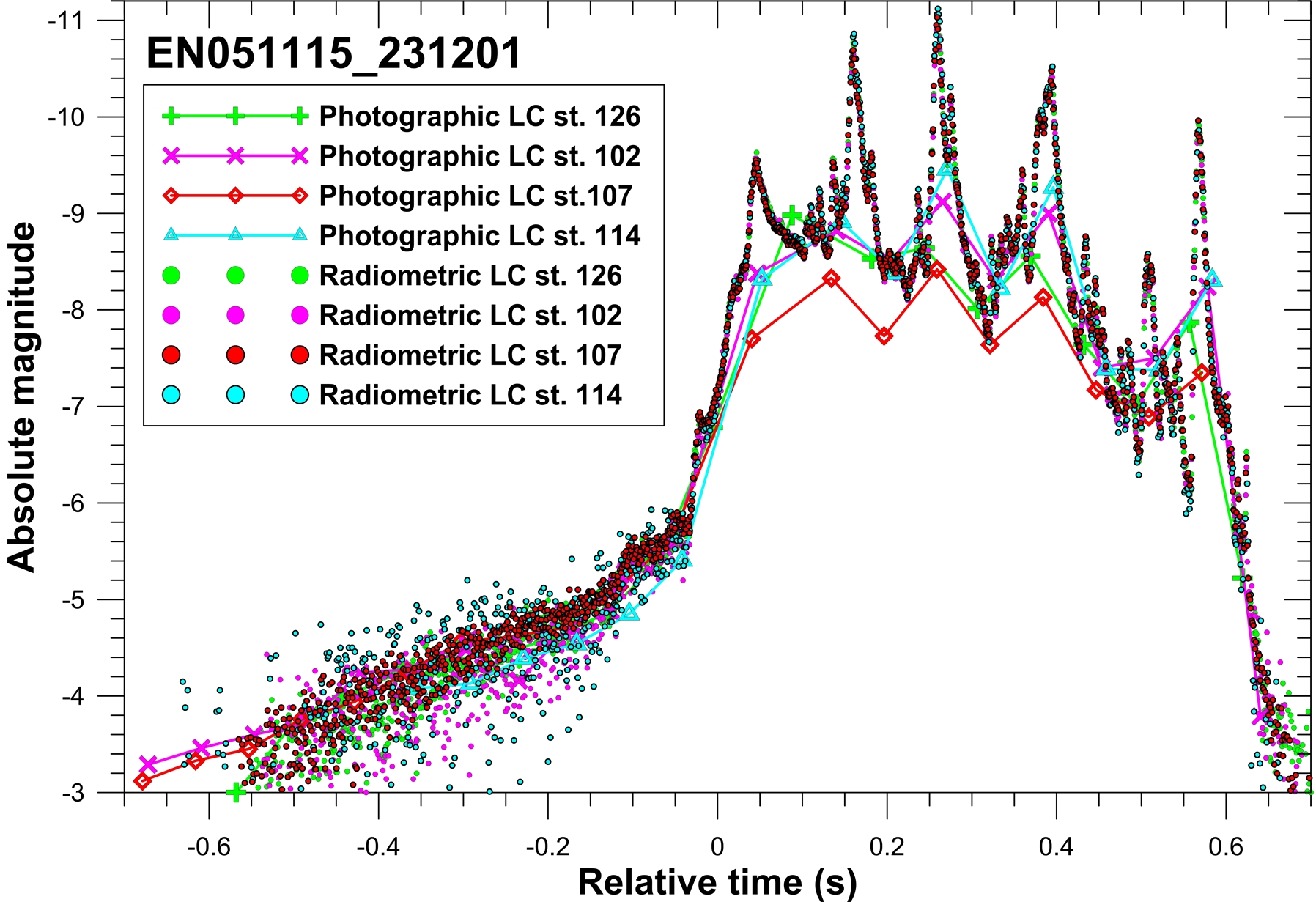}
\caption{Photographic and radiometric light curves of the EN051115\_231201 fireball in absolute magnitudes.}
\label{lcmag}
\end{figure}

The first step after measurement of all four digital images and their astrometric reduction is computation of the atmospheric luminous trajectory.
We use two different methods described in \citet{Cep87}: the so-called plane method, and in \citet{Bor90}, the so-called least-squares method.
A first independent check of the results is that the values describing the atmospheric trajectory obtained
from these two methods turn out to be the same within the uncertainties.  Lateral deviations of all measured points from the resulting atmospheric trajectory (zero line) are shown
 in Figure~\ref{trajectory}. This plot illustrates the high reliability of the astrometric solution. The spread of the measured points
from individual stations is random and the standard deviation is only 7 m. In this context it is also important to mention 
how far each station (camera) was from the fireball. Exact distances of the beginning and terminal points $R_{(B \div E)}$ 
for each station were as follows:

\vspace{2.5mm}

\noindent $R_{(B \div E)} \; (107) = 141.8 \div \; \: 92.7$ km; \\
\vspace{-2.5mm}

\noindent $R_{(B \div E)} \; (126) = 149.1 \div 119.5$ km; \\
\vspace{-2.5mm}

\noindent $R_{(B \div E)} \; (102) = 176.6 \div 143.3$ km; \\
\vspace{-2.5mm}

\noindent $R_{(B \div E)} \; (114) = 221.5 \div 189.8$ km. \\

This example nicely illustrates that our records and methods provide us with a precision of the atmospheric trajectory 
determination of about 10 m for fireballs that are still about 200 km away the stations. Most of the Taurids in this study, 
especially the fainter ones, were below or around this distance, and only several of the brightest cases were at much larger distances 
from the stations. The most distant Taurid was the superbolide EN311015\_180520, which was also recorded by the cameras 
at distances of up to 630 km (stations that were used for analysis). Data precision for such a distant and difficult to measure 
case is about 140 meters, which is still good.
The precision is crucial not only for the determination of the position of the trajectory in the atmosphere, but also
for the determination of the direction of flight of the meteoroid, in other words, the position of the apparent radiant, which is important
for determination of the heliocentric orbit of the meteoroid. The coordinates of the apparent radiant for this particular fireball were
\vspace{2.5mm}

\noindent$\alpha_{\rm app} = 54.947\degr \pm 0.007\degr$, $\delta_{\rm app} = 16.196\degr \pm 0.004\degr$. \\

When computing the local azimuth and slope of the trajectory, we took into account the curvature of bolide trajectory due to gravity,
 which can be significant for longer fireballs with very precise data. For the EN051115\_231201 it is only 0.02\degr.

The second step is the determination of the velocity of the fireball. The data in this study are so good that it enabled us
to use the method described in \citet{Cep93} for the vast majority of fireballs. Successful application of this model is very sensitive
to the quality of the lengths for each individual measured velocity point corresponding to a single shutter break.
This rigorous physical model provides the speed at any point on the trajectory but for the presented study, which is focused
on the orbital analysis, the initial velocity is the most important.  For the sample fireball we obtained a four\ parameter (non-fragmenting)
solution including initial velocity for each station and all solutions were very similar. Nevertheless for the final dynamic 
solution and initial velocity determination, not only for this particular case but for all cases in this study, we used a slightly different approach. 
We put all measured shutter breaks from all used images together (timescales on all DAFOs are correlated)
and we applied the Ceplecha method on this unified data set. This approach significantly increases reliability
of the resulted dynamic solution. It is useful especially for shorter fireballs such as Taurids because they are moderately fast meteoroids
of cometary origin (i.e., relatively fragile) and the number of measured breaks on one image can be limited. Therefore, 
every independent measurement can be very useful in obtaining a reliable value of the initial velocity.
Moreover, for some fireballs the non-fragmenting solution applied to the whole trajectory was not adequate and we had to omit the terminal
part of the fireball to obtain a realistic value of the initial velocity. This was also the case of the sample Taurid fireball EN051115\_231201 
as can be seen for example in Figure~\ref{twolcint} where several bright flares corresponding to fragmentation events are clearly visible. 
The resulting value of initial velocity of the EN051115\_231201 Taurid fireball is $31.221 \pm 0.037$ km s$^{-1}$.

\begin{figure*}
\centering
\includegraphics[width=0.9\hsize]{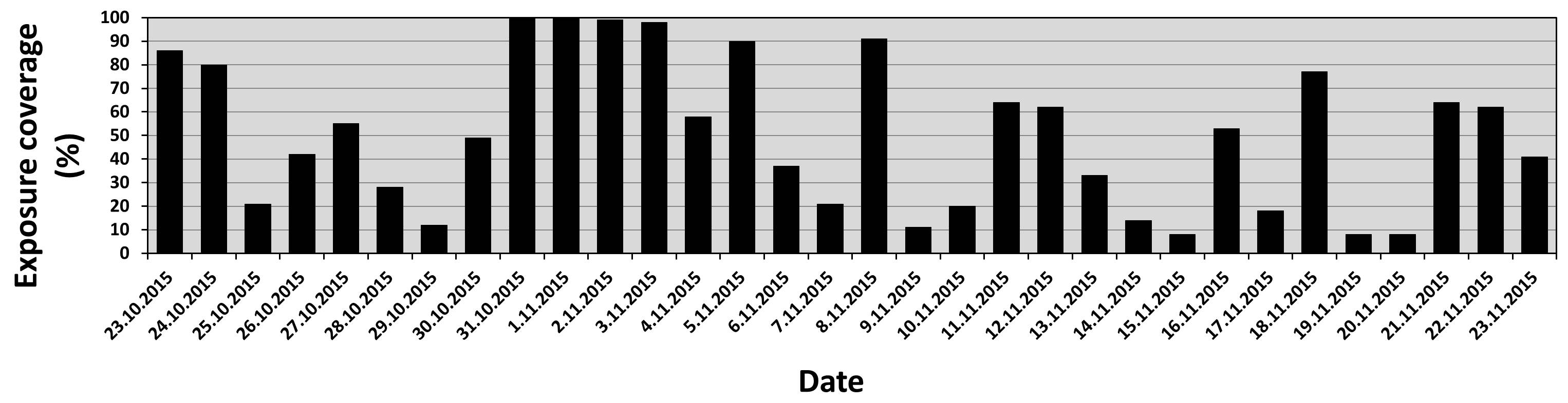}
\caption{Exposure coverage representing observing conditions in Central Europe during activity of Taurids in 2015.
It is the fraction of real time when all cameras in the network exposed to their total prescribed exposure time. The date corresponds to the evening date of the whole night.}
\label{expcoverage}
\end{figure*}

\begin{figure*}
\centering
\includegraphics[width=0.9\hsize]{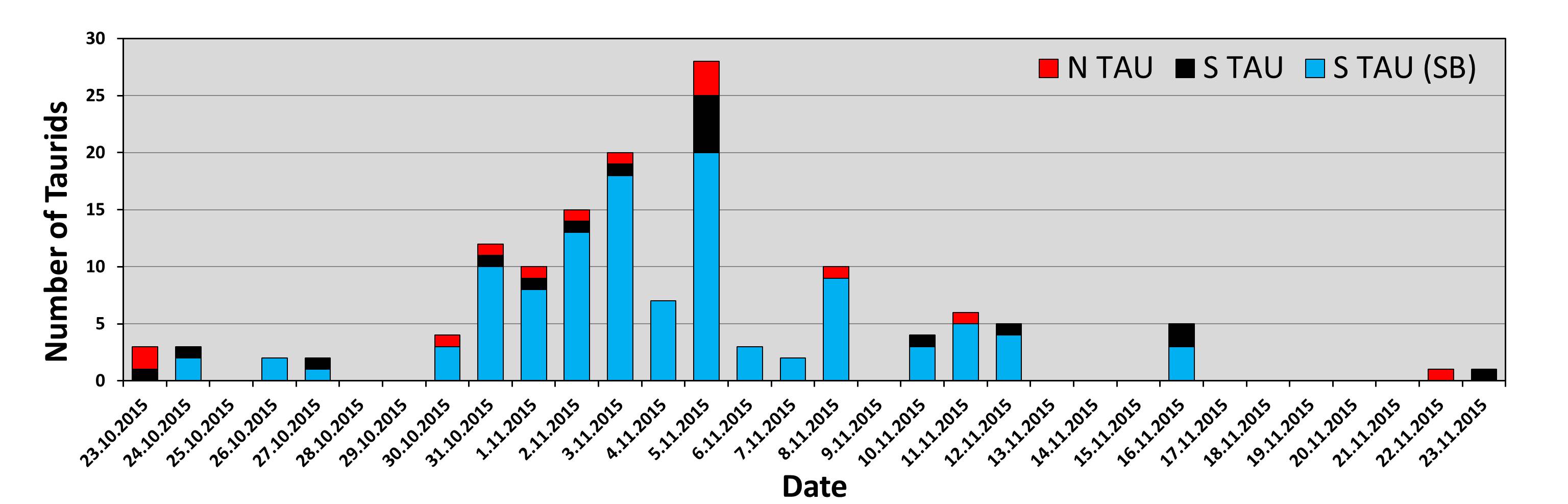}
\caption{Activity of Taurids recorded by the DAFO cameras of the European Fireball Network in 2015. The number of presented
Taurid fireballs in the plot is 143 (total number is 144); S TAU from 28.11.2015 is out of range of the date axis. Date corresponds to the evening date of the whole night.}
\label{activity}
\end{figure*}

The next step in the analysis of the available records is the exact photometry of the fireball. We have two different data types, those from photographic records and radiometers, from which
we can determine the brightness of the fireball and its initial mass based on photometry.
As mentioned above, we measure digital images in 14-bit raw format. We found that this limited dynamic range
of the used CMOS sensor is sufficient for fireballs with apparent magnitude up to about $-8$. Above this limit the measured signal
starts to be saturated. As shown in Table~\ref{tbphysical}, which contains basic physical data of the presented Taurids,
this method can be reliably used for about 75\% of all cases. The remaining 25\% of presented cases are
such bright fireballs that their digital images are partly or even almost completely saturated. For such fireballs we have 
different methods to describe their brightness. One solution to this problem is the use of simultaneous photographic images
taken by the AFO on the film. The response of the film emulsion is logarithmic, which means that the photographic film has
much higher dynamic range; we use Ilford FP4 panchromatic films with sensitivity 125 ASA. This is a quite straightforward method
and we used it in few cases, but a still much more appropriate and accurate way is the use of the light curves taken by the radiometers,
which are in our cameras and still have much higher dynamic range. The apparent (i.e., not corrected for distance)
high-resolution (5000 samples/s) radiometric light curves of the EN051115\_231201 fireball taken by radiometers
at Kun\v zak (blue) and \v Cerven\' a hora (red) stations are shown in Figure~\ref{twolcint}. The close agreement between the different records, a testament to the high precision of the data, is evident. As for other two closer stations to the fireball, the radiometric light curves
have exactly the same profile and could be used for fireball photometry. However, we cannot use radiometric light curves
 directly because individual radiometers have different sensitivity and are not calibrated to obtain absolute photometry.
For the purpose of calibrating these records we combine photometry from both methods. We measure
the meteor signal on the digital image and for calibration we use that part of the photographic image, which is not saturated and at the same time well above the noise of the measured signal from both the photographic records and corresponding radiometric
records. This is usually somewhere in the interval between $-4$ and $-7$ magnitude. However, we have to relate the timescale
of the photograph to the absolute timescale of the radiometer. For this purpose we use time marks (breaks of double length) made by the electronic shutter along the luminous path of the recorded fireball on the beginning of each second. This defines the exact absolute time
of this measured point, which can be simply identified with the corresponding point on the radiometric light curve.
Both radiometer and electronic shutter are continuously corrected by the PPS pulse of the GPS so the absolute timing of both
records is given with high precision.

The result of this procedure is illustrated on Figure~\ref{lcmag} in which photographic and radiometric light curves from
all used stations in absolute magnitudes are plotted. The first evident result is that, especially for shorter and faster fireballs
containing bright and short significant flares, the photographic photometry cannot correctly describe the shape of the light curve because
of the low time resolution. The electronic shutter, on the other hand, has a resolution of 16 interruptions per second with the same length
for the on/off state. This means that blind time lasts exactly 0.03125 seconds and it is sometimes longer than the duration
of a flare or at least its brightest part. As can be seen in Figure~\ref{twolcint} and Figure~\ref{lcmag}, this is exactly
the case of fireball EN051115\_231201, where most of the light is contained within five distinct and quite short flares
that are only partly recorded (or even missing) on the image. Another aspect that is evident in Figure~\ref{lcmag}
is the saturation of the photographic records.  The absolute photographic photometry profile (brightest parts of the light curves)
is, unlike the radiometric photometry, quite different for individual stations. However it confirms the saturation effect
because stations, which are more distant from the fireball give higher absolute maximum brightness. It means that these records
are not as saturated as the records from closer stations. The last aspect, which is worth mentioning in connection with the photometry
shown in Figure~\ref{lcmag}, is that for correct calibration of radiometric light curves it is much better to use the radiometric record
and photographic image from the closest station where the signal-to-noise ratio is the most favorable. This is valid especially for the fireballs
as in the case described here, when the increase of the brightness is very steep and the suitable (not saturated) interval of magnitudes is very short.

A general conclusion is that the high-resolution radiometric records are crucial for correct recovery
of the photometry of all brighter fireballs, especially those that contain distinct flares. We note that the photometry based on radiometric
light curves was determined for about 90\% of analyzed Taurids in this study.

As explained above, apart from the precise photometry of the recorded fireballs, radiometric records provide us with a very accurate
absolute time of each event. This important parameter is, along with the initial velocity and radiant position,
necessary for reliable determination of the heliocentric orbit of the observed fireball. The orbits were computed by the method of
\citet{Cep87}.

To compute the photometric mass of the meteoroid, the velocity dependence of the luminous efficiency was taken from \citet{ReVelle2001}.
The mass dependence was ignored by substituting 10 kg for the mass in their formula. Specifically, the luminous efficiency for velocities
above 25.4 km s$^{-1}$ was assumed to be directly proportional to the velocity, reaching 6.5\% at 30 km s$^{-1}$.
A factor of 1500 W for zero magnitude meteor \citep{SSR} was used to convert magnitudes into
bolometrically radiated  energy.

The above-mentioned example clearly demonstrates high precision and reliability of all parameters describing
the atmospheric trajectory, dynamics, photometry, and heliocentric orbit not only for this particular case, but also for all Taurid fireballs presented in Table~\ref{tborbits} and Table~\ref{tbphysical}. For this complex analysis of all presented
Taurid fireballs, i.e., the astrometric reduction of the images, atmospheric trajectory computation, dynamic and photometric solutions, and
finally orbital calculations, we used our new software package BOLTRACK (J. Borovi\v cka).

Although the autumn weather, especially in November, is notoriously cloudy in Central Europe, the year 2015 was not so bad.
There were several clear nights, especially in the beginning of November, and only a few nights were completely cloudy practically at all
stations. This situation is illustrated in Figure~\ref{expcoverage} in which the ratio of real to prescribed exposure time for all stations
in the network altogether and for each individual night covering the Taurids activity in 2015 is shown. Since the DAFOs also work when the sky is only partly clear, some fireballs were captured even in the nights when it was mostly cloudy and
could be still used for this study. On the other hand, some fireballs were recorded only from one station or their records were of a quality that is insufficient to merit scientific analysis. Such cases were excluded from our study. As a result of relatively favorable weather conditions and  the capability
of our network, we were able to cover the whole period of the enhanced Taurid activity from the last decade of October to mid-November as shown in Figure~\ref{activity}. It is difficult to construct the activity profile for such a long interval
from our data because it is difficult to take into account all observational effects and correctly eliminate them. So Figure~\ref{activity} does not represent the real activity profile, but only the uncorrected distribution of selected
Taurid fireballs during the whole interval of activity. As described in the following sections,
we identified three different groups of Taurids in our data set. These are the regular Southern  and Northern  Taurids (designated as
S TAU and N TAU, respectively) and a new branch of Southern Taurids, which we designate S TAU (SB). As shown later, this new branch was responsible for the enhanced activity. From Figure~\ref{activity} we can see that enhanced Taurid activity caused by this new branch of Southern Taurids fireballs started on October 24, culminated around November 5 and terminated
on the night of November 16/17. The S TAU (SB) was not observed after this date even though on November 18 and 21-23 were at least partly clear nights
with good observing conditions. From Figure~\ref{activity}
we also see that the enhanced activity increased gradually with several days of very high activity at the turn of October and November.
This interval was strongly affected by the full Moon period, so fainter fireballs were
below the sensitivity limit of the digital all-sky cameras and radiometers especially in the end of October. Therefore the number
of fireballs on these nights may be underestimated.
On the other hand, the relatively steep decrease of activity after November 5 seems to be real.
With regard to regular S and N Taurids, they are quite uniformly spread over the entire interval of observed activity.

\begin{figure*}
\centering
\includegraphics[width=0.9\hsize]{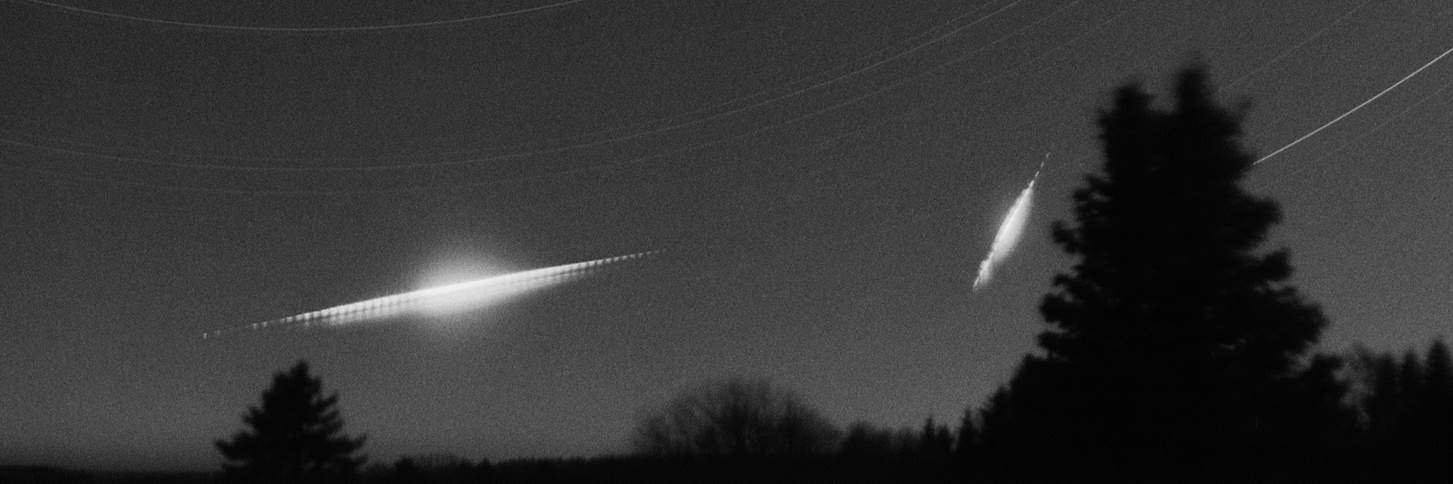}
\caption{Detailed view of the two brightest Taurids far over Poland recorded by the AFO (analog camera) at station Polom.}
\label{twotaurids}
\end{figure*}

\section{The 2015 Taurid data}
\label{data}

The total number of Taurid fireballs recorded photographically by our instruments at least from two stations in 2015 was about 200.
This is much more than we recorded in any previous year. The main reason for this is evidently the unusually high Taurid activity,
but it is also caused by much more efficient observational system and also by a quite long
period of relatively good weather. For this study we selected 144 Taurids with complete information about heliocentric orbits of individual meteoroids and their physical properties as well. Our data set is unique not because of the total number of used meteors but
the high precision of the data for each individual case, which was obtained by with high-resolution cameras
and radiometers and elaborated reduction methods. Meteor records were reduced one-by-one and
all steps in the measurement and computation process were under careful human supervision.

Before going to the statistical analysis of the whole data set, we describe in more detail some remarkable fireballs.

\subsection{Exceptional cases}

\subsubsection{EN311015\_180520 and EN311015\_231301 - two brightest Taurids}
\label{twobrtau}

It is a well-known fact that Taurids are quite rich in bright fireballs. However, 2015 was also in this aspect exceptional
and our cameras recorded several very bright Taurids during the whole period of activity. Altogether 24 Taurids were brighter
 than $-10$ absolute magnitude and 10 were similarly bright or even brighter than the full Moon. Moreover, two of those, both
observed in the first half of the night of October 31, are really remarkable not only in this data set, but in all Taurids that we have recorded within the EN until now. Both are shown in Figure~\ref{twotaurids}, where a small
part of the all-sky image is shown. This image was taken from station Polom by the AFO, i.e.,\ on sheet film, where one
exposure was taken per night. Both bolides were observed in a similar (northern) direction and flew over northern and central Poland, respectively.
The brighter bolide, which reached a peak absolute magnitude of $-18.6$ (on the left), occurred at 18:05:20 UT and the second,
with peak magnitude of $-15.8,$ occurred 5 hours 7 minutes and 41 seconds later at 23:13:01 UT. This is the reason why their
directions of flight differ, although both bolides had practically the same radiant. The first was
so bright that it belonged to the superbolide category. This spectacular Taurid bolide was caused by a meteoroid
with initial mass more than 1000 kg, i.e.,\ a meter-sized object. Because of its enormous
brightness, clear skies over large parts of Central Europe, and convenient time of its occurrence (it was an unusually nice Saturday evening),
thousands of eyewitnesses were fascinated by this extraordinary natural event. We obtained more reports of
one bolide than ever before. Apart from plenty of visual observations, all DAFOs and AFOs in our network (at 15 stations) recorded it, which was crucial for reliable description of this superbolide. In addition to our own
photographic and radiometric records, we used also two casual images.
The first one is a high-resolution digital image from Stud\' enka, Czech Republic, which was obtained from amateur
astronomer B. Pelc. The second digital image was taken by G. Zieleniecki at Czernice Borowe, Poland, and was freely
available on the internet. Altogether we used 13 most suitable photographic and
5 radiometric records. The situation with the second, much smaller,  meteoroid was similar. It was also
recorded by all our cameras at all 15 stations, and we obtained also 4 high-resolution casual digital images from northern
region of the Czech Republic. These images, which we also partially used, were taken by T. Chl\'{\i}bec at Kl\'{\i}novec,
L. Sklen\' ar from Kun\v cice and Labem, D. \v S\v cerba from Doln\'{\i} \' Udol\'{\i}, and L. Shrben\' y from \v R\'{\i}\v cany;
this record also includes a spectrum of the bolide. In this case we used the best 12 photographic and 4 radiometric
records for final analysis. 

Since these fireballs were exceptional,
we modeled them with our semiempirical fragmentation model \citep{Kosice}. The model fits radiometric curves
and deceleration. This way we obtained more reliable initial masses of meteoroids (1300 kg and 34 kg, respectively)
and insight into their atmospheric fragmentation. Both meteoroids were effectively destroyed high in the atmosphere under dynamic
pressures $<0.05$ MPa. In both cases a small fragment ($< 1$ kg) survived the initial destruction and fragmented further
under pressures of $\sim$ 0.1 MPa. In comparison with other bright bolides \citep{Romanian}, both Taurids were extremely fragile.

\begin{table*}
\caption{Atmospheric trajectory data for the EN311015\_180520 (left) and EN311015\_231301 (right) bolides.}
\label{twobolatm}
{\normalsize
\begin{tabular}{@{\extracolsep{\fill}} l|r@{ $\pm$ }lcr@{ $\pm$ }l|r@{ $\pm$ }lcr@{ $\pm$ }l}
\hline \\ [-3mm]
& \multicolumn{2}{c}{Beginning} & Max. & \multicolumn{2}{c|}{Terminal} &
 \multicolumn{2}{|c}{Beginning} & Max. & \multicolumn{2}{c}{Terminal} \\
& \multicolumn{2}{c}{} & bright. & \multicolumn{2}{c|}{} & \multicolumn{2}{|c}{} & bright. & \multicolumn{2}{c}{} \\ [0.1mm]
\hline \\ [-3mm]
Height (km)	& 114.724 & 0.025 & 80.8 & 57.644 & 0.030 		& 120.026 & 0.030 & 74.4 & 57.305 & 0.016 \\
Velocity (km s$^{-1}$)	& 33.07 & 0.03	& 33.07 & 22 & 2 			& 32.56 & 0.09	& 32.53 & 30 & 2	\\
Longitude ($\degr$ E)	& 18.46416 & 0.00018	& 16.927 & 15.82901 & 0.00020 & 18.18064 & 0.00026 & 18.101 & 18.07087 & 0.00014 \\
Latitude ($\degr$ N)	& 53.60723 & 0.00050	& 53.553 & 53.50244 & 0.00048 & 52.13173 & 0.00060 & 52.430 & 52.54390 & 0.00033 \\
Slope ($\degr$)	& 18.574 & 0.014 & 17.71 & 17.148 & 0.034 &	53.28 & 0.04 &	53.00	& 52.89 & 0.10	\\
Azimuth ($\degr$)	& 267.240 & 0.013 &	266.00	& 265.120 & 0.015 &	350.78 & 0.04	& 350.72 & 350.69 & 0.05	\\
Time\footnotemark[1](s)
 & \multicolumn{2}{c}{$-1.07$} & 2.22 & \multicolumn{2}{c|}{4.68} & \multicolumn{2}{c}{$-0.07$} & 1.68 & \multicolumn{2}{c}{2.36}
\\[0.5ex]
Total length (km) & \multicolumn{2}{c}{} & 186.3 & \multicolumn{2}{c}{}  & \multicolumn{2}{|c}{} & 78.5 & \multicolumn{2}{c}{} \\[0.05mm]
\hline
\end{tabular}
}
\footnotemark[1]{\tiny Time zero corresponds to 18:05:18 UT for EN311015\_180520 and
23:13:00 UT for EN311015\_231301.}
\end{table*}

\begin{table}
\caption{Apparent and geocentric radiants and orbital elements (J2000.0) for the EN311015\_180520 (left) and EN311015\_231301 (right) meteoroids.
Time is given for the average point of the recorded trajectory.}
\centering
\label{twobolorb}
{
\begin{tabular}{@{\extracolsep{\fill}} r@{ }l|r@{ $\pm$ }l|r@{ $\pm$ }l}
\hline \\ [-3mm]
&& \multicolumn{2}{|c}{EN311015\_180520} & \multicolumn{2}{|c}{EN311015\_231301} \\
\hline \\ [-3mm]
Time	&	(UT)	&	\multicolumn{2}{c|}{$18^h05^m20.0^s$ $\pm$ $0.1^s$} & \multicolumn{2}{|c}{$23^h13^m01.5^s$ $\pm$ $0.1^s$}	\\
$\alpha _R$	&	($\degr$)	&	50.126 & 0.009 & 51.853 & 0.022	\\
$\delta _R$	&	($\degr$)	&	16.452 & 0.016 & 15.66 & 0.04	\\
v$_{\infty }$	&	(km s$^{-1}$)	&	33.068 & 0.030 & 32.56 & 0.09	\\
$\alpha _G$	&	($\degr$)	&	51.692 & 0.010 & 51.445 & 0.022	\\
$\delta _G$	&	($\degr$)	&	14.592 & 0.017 & 14.49 & 0.04	\\
v$_G$	&	(km s$^{-1}$)	&	30.869 & 0.032 & 30.59 & 0.10 \\
v$_H$	&	(km s$^{-1}$)	&	37.32 & 0.02 & 37.34 & 0.06 \\
$a$	&	(AU)	&	2.250 & 0.009 &	2.258 & 0.027 \\
$e$	&			&	0.8724 & 0.0006 & 0.8689 & 0.0020 \\
$q$	&	(AU)	&	0.28715 & 0.00032 & 0.2960 & 0.0010 \\
$Q$	&	(AU)	&	4.212 & 0.018 &	4.22 & 0.05 \\
$\omega$ 	&	($\degr$)	&	121.687 & 0.022 &	120.62 & 0.06 \\
$\Omega$ 	&	($\degr$)	&	\multicolumn{2}{c|}{37.791} & \multicolumn{2}{c}{38.005}	\\
$i$	&	($\degr$)	&	5.707 & 0.023 & 5.62 & 0.06  \\
$P$	&	(yr)	&	3.375 & 0.020 &	3.39 & 0.06  \\
 \multicolumn{2}{l|}{Perihelion} & 2012-07-26 & 7 d  & 2012-07-20 & 22 d\\
\multicolumn{2}{c|}{TP$_{\rm Jup}$}	& 2.952 & 0.009 & 2.953 & 0.028	\\
\hline\noalign{\smallskip}
\end{tabular}
}
\end{table}

\begin{figure}
\centering
\includegraphics[width=0.9\hsize]{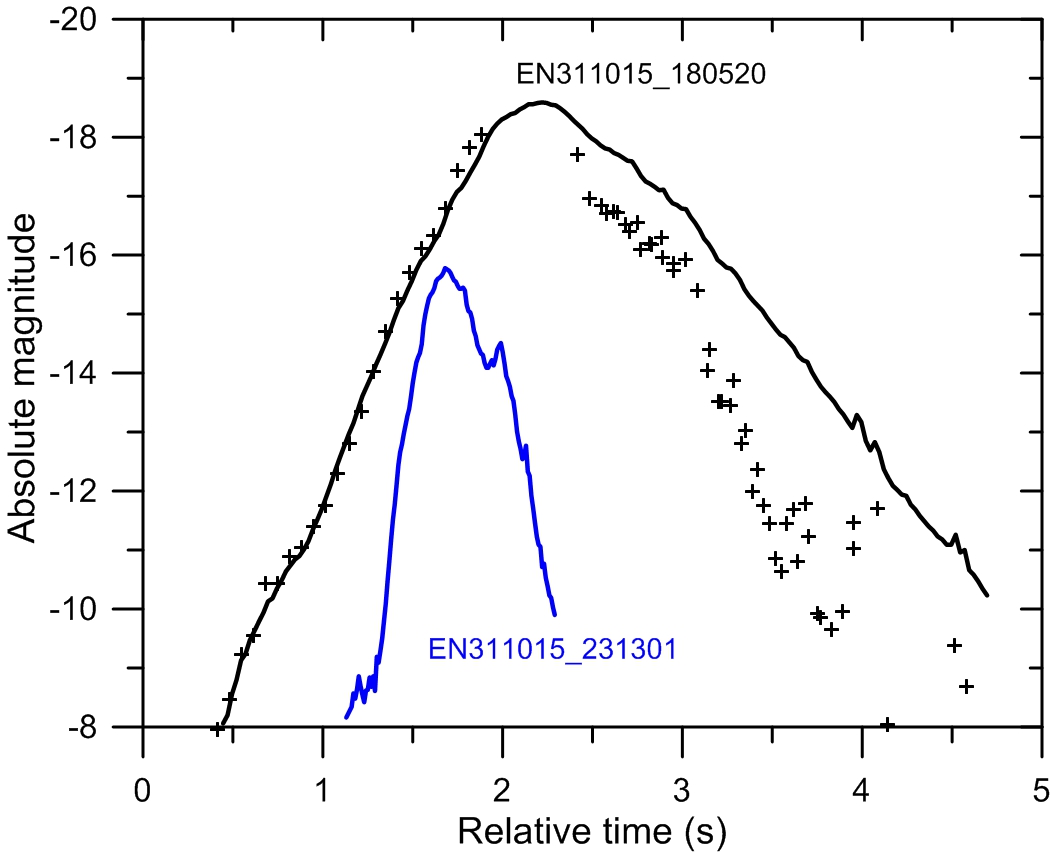}
\caption{Calibrated radiometric light curves  of bolides EN311015\_180520 and EN311015\_231301 (solid curves).
For EN311015\_180520 data from two imaging cameras are also given  (crosses).
After bolide maximum, most of radiometric signal was produced by a stationary
trail. Camera data contain only the bolide moving further down. Time zero corresponds to 18:05:18 UT for EN311015\_180520 and
23:13:00 UT for EN311015\_231301.}
\label{twobollc}
\end{figure}

Simultaneously with our network, both bolides were also recorded by the cameras of the Polish Fireball Network. These data were
analyzed independently and were published by \citet{Olech}. Because our data differ from their data,
we provide here our complete results and compare them to those reported in \citet{Olech}.  Atmospheric trajectories are given in Table~\ref{twobolatm},
light curves in Fig.~\ref{twobollc}, and  heliocentric orbits in Table~\ref{twobolorb}. When computing the local
azimuth and slope of the trajectory, we took into account both the curvature of the Earth and the curvature of the bolide trajectory
due to gravity, which was significant for EN311015\_180520 (change of direction of flight by 0.17$\degr$ over the recorded length).
Azimuths are measured from the south clockwise. The apparent radiants given in Table~\ref{twobolorb} are valid for the average points on
the trajectories.

As shown later in this paper and that of \citet{Olech} (in fact that paper is based only on
these two bolides) data about these big bolides are of great importance. Since it may not be simple to distinguish which data set is correct, we carry out an analysis of the differences.
The positions of the recorded beginning and end points of the bolide depend on the sensitivity of the instrument and the observing conditions.
Nevertheless, when plotting the trajectory of the first bolide on the map, the solution of \citet{Olech} is shifted about 1.8 km to the north.
As for the observed apparent radiant, there is a difference of 0.65$\degr$ in declination, i.e.,\ 10 times their quoted uncertainty ($\sigma$).
For the second bolide, the larger difference is in right ascension (0.24$\degr$, $4\sigma$) and especially in entry velocity, which is larger by 0.6 km s$^{-1}$ ($6\sigma$) in \citet{Olech}.
Although our data were obtained from large distances, our results are based on large
number of records (in both cases more than 10) and the solutions for both bolides are very consistent. The photograph
from Czernice Borowe, when combined with our cameras, provides convergence angles in excess of 60$\degr$ for the first bolide.
Czech cameras have mutual convergence angles up to 15$\degr$. For the second bolide the situation is even better,
although the velocity was more difficult to measure.
So there is no reason, why our results should be different by more than the standard deviations given in Table~\ref{twobolatm}.

There is also at least a 3-5 seconds difference in the reported time of appearance of the first fireball.
Our radiometers are continuously corrected by PPS pulse of GPS and  their timing precision is in millisecond range.
A nice example how our radiometers are synchronized is shown in Figure~\ref{twolcint}.
Another discrepancy is in the determination of the maximum
absolute brightness for both bolides.
While \citet{Olech} determined the maximum absolute brightness $-16.0\pm0.4$ mag for the first bolide, we found that it reached
$-18.6 \pm0.2$ mag.  Our method described in Section~\ref{precision} relies on the linearity of radiometers even for strong signals.
Results from five independent radiometers were in perfect agreement.
Similarly the brightness of the second bolide was underestimated by \citet{Olech} by about 1 magnitude.

\begin{figure*}
\centering
\includegraphics[width=0.9\hsize]{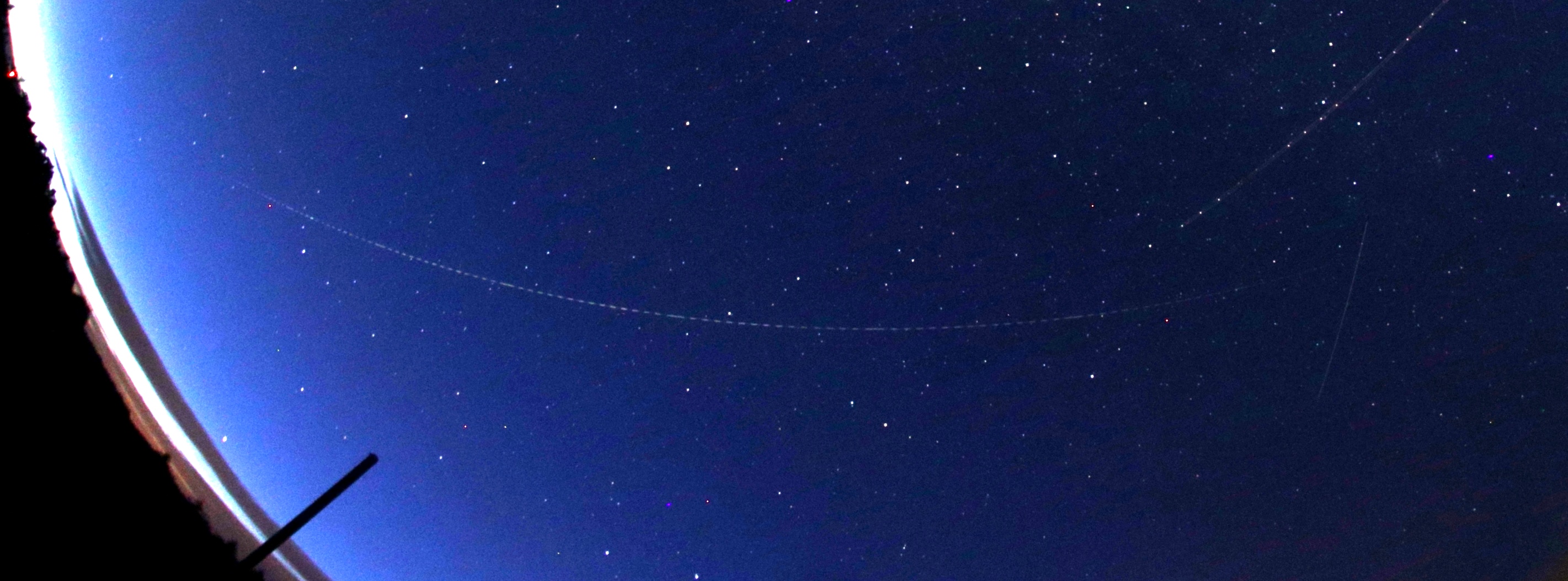}
\caption{Detailed view on the very long Taurid fireball recorded by the DAFO (digital camera) at station Kun\v zak.}
\label{longtaurid}
\end{figure*}

Regardless of these differences in the directly determined trajectory parameters, we found another 
discrepancy in the calculation of orbital elements in Tables 3-5 of \citet{Olech}.
We obtained significantly different results than those published
in Tables 5 and 6 of \citet{Olech} when we took their input values of initial velocity, apparent radiant position (which  we recalculated  from J2000.0 to the date that the bolides occurred) time, and mean position for both bolides from their Tables 3 and 4 and used our program for orbital calculation; orbits from
this program were independently validated for example in \citet{Clark}. We found the following differences for the EN311015\_180520 bolide (computed minus published):
$\Delta a$ = 0.0077 AU, $\Delta e$ = 0.0057, $\Delta\omega$ = 1.3$\degr$ (!), $\Delta i$ = 0.07$\degr$, and $\Delta P$ = 0.019 yr.
For the EN311015\_231301 bolide, differences are
$\Delta a$ = 0.1287 AU (!), $\Delta e$ = 0.0087, $\Delta\omega$ = 0.1$\degr$, $\Delta i$ = 0.035$\degr$, and $\Delta P$ = 0.20 yr.
Some of these differences are really high, namely 1.3$\degr$ , in argument of perihelion for the first bolide and
especially 0.128 AU in semimajor axis for the second bolide. With the radiant and velocity given by \citet{Olech}
this bolide would be far from the 7:2 resonance with Jupiter, nevertheless, their published orbit puts it in the resonance.

\subsubsection{EN061115\_164758:\ An almost horizontal Taurid}

On November 6, 2015 during dawn, just after the Taurid radiant rose above the horizon, a relatively faint Taurid fireball of $-5.1$
maximum absolute magnitude traveled over a large part of sky and was observed by several stations in the SW part of our network. The sky was
not completely dark, especially from the stations in western part of the network, which were closest to the fireball trajectory.
However, thanks to the higher sensitivity of the digital cameras, this extremely long fireball was nicely recorded on three stations, Kun\v zak, 
Martinsberg and Kucha\v rovice, which enabled us to describe this exceptional Taurid accurately. Owing to its small slope which was  7.7$\degr$ 
at the beginning
and during the flight decreased to only 5.6$\degr$, the recorded fireball trajectory was extremely long, i.e., exactly 258.7 km, and its
flight lasted 8.5 seconds. It is the longest Taurid fireball we have ever recorded, both in duration and length.
Thanks to a large amount of data points, this
Taurid has the best dynamic data and the trajectory, i.e., also radiant, is also very precise. The initial velocity of 31.285 km s$^{-1}$ was 
determined with a precision of $\pm$7 m s$^{-1}$. As for the brightest Taurid described in Sect.~\ref{twobrtau}, 
we took into account the curvature of the trajectory of the bolide due to gravity, which was significant for such a long and low inclined fireball (change of direction of flight by 0.18$\degr$ 
over the recorded length).
A\ detailed view of its luminous flight taken by the DAFO at Kun\v zak station is shown in Figure~\ref{longtaurid}.
Additional information about this fireball is given in Tables~\ref{tborbits} and \ref{tbphysical}.

\subsection{Radiants and orbits}

\begin{figure}
\centering
\includegraphics[width=0.79\hsize]{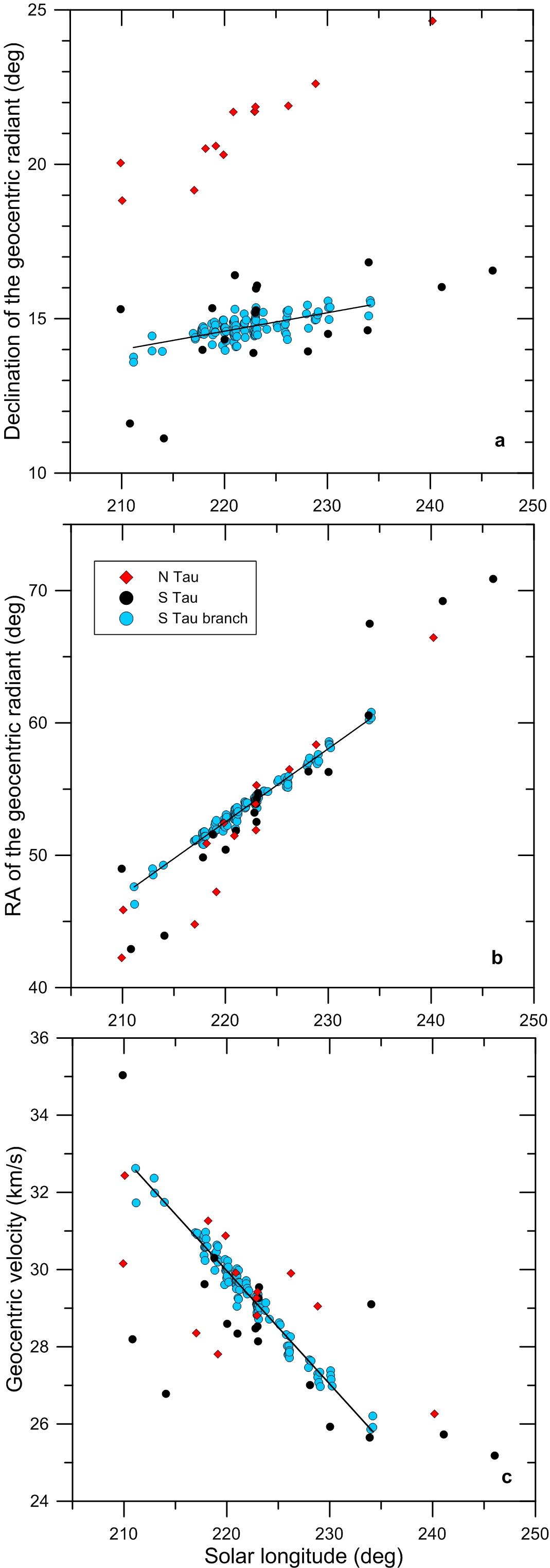}
\caption{Position of geocentric radiant and geocentric velocity as a function of solar longitude for Taurids observed in 2015.
Northern and Southern Taurids are shown by different symbols and the Southern Taurids belonging to the new branch are highlighted
in light blue. The error of the data is smaller than the size of the symbols in most cases.}
\label{radiant}
\end{figure}

In this section the radiants, velocities, and heliocentric orbits of all 144 fireballs are evaluated.
All elements in this paper are given for equinox J2000.0. The data are presented in Table~\ref{tborbits}.
Figure~\ref{radiant} shows the dependency of geocentric radiant and velocity on solar longitude (i.e.,\ the longitude of the Sun
at the time of fireball observation). Thirteen fireballs were classified
as Northern Taurids. They can be easily recognized by their radiant lying to the north of the ecliptic. All other fireballs
belong to Southern Taurids. Among them, a well-defined structure can be recognized, where the radiant position
and velocity are strict linear functions of solar longitude. We call this structure a new branch.
Evidently, this branch was responsible for the enhanced Taurid activity in 2015.

Regular Taurids also exhibit radiant motion but the spread of individual radiants is much larger than
for the new branch. For the new branch, we found the following relationships:
\begin{eqnarray}
\alpha_{\rm g} & = & 46.99\degr + 0.554\cdot (\lambda_{\sun} - 210\degr) \\
\delta_{\rm g} & = & 14.00\degr + 0.060\cdot (\lambda_{\sun} - 210\degr) \\
v_{\rm g} & = & 32.90 - 0.293\cdot (\lambda_{\sun} - 210\degr),
\end{eqnarray}
where $\alpha_{\rm g}$ and $\delta_{\rm g}$ are the right ascension and declination, respectively, of the geocentric radiant
(J2000.0), $v_{\rm g}$ is the geocentric velocity in km s$^{-1}$, and $\lambda_{\sun}$ is solar longitude  (J2000.0).
Although we defined the new branch on the basis of orbital elements rather than radiants and velocities (see below), all fireballs
of the new branch had the radiant right ascension within $1.3\degr$ and declination within $0.7\degr$ from (1) \& (2).
The velocities were within
0.9 km s$^{-1}$ from relationship (3). We point out that this spread is real. The precision of most of our data, as demonstrated
in Sect.~\ref{precision},  is < $0.05\degr$ in the radiant position and 0.1 km s$^{-1}$ in the velocity.

Fireballs from the branch were observed between
solar longitudes $211\degr$ -- $234\degr$ (October 25 -- November 17).
Taurids observed before and after these dates belong to the background population of Northern and Southern Taurid streams.

\begin{figure*}
\centering
\includegraphics[width=0.9\hsize]{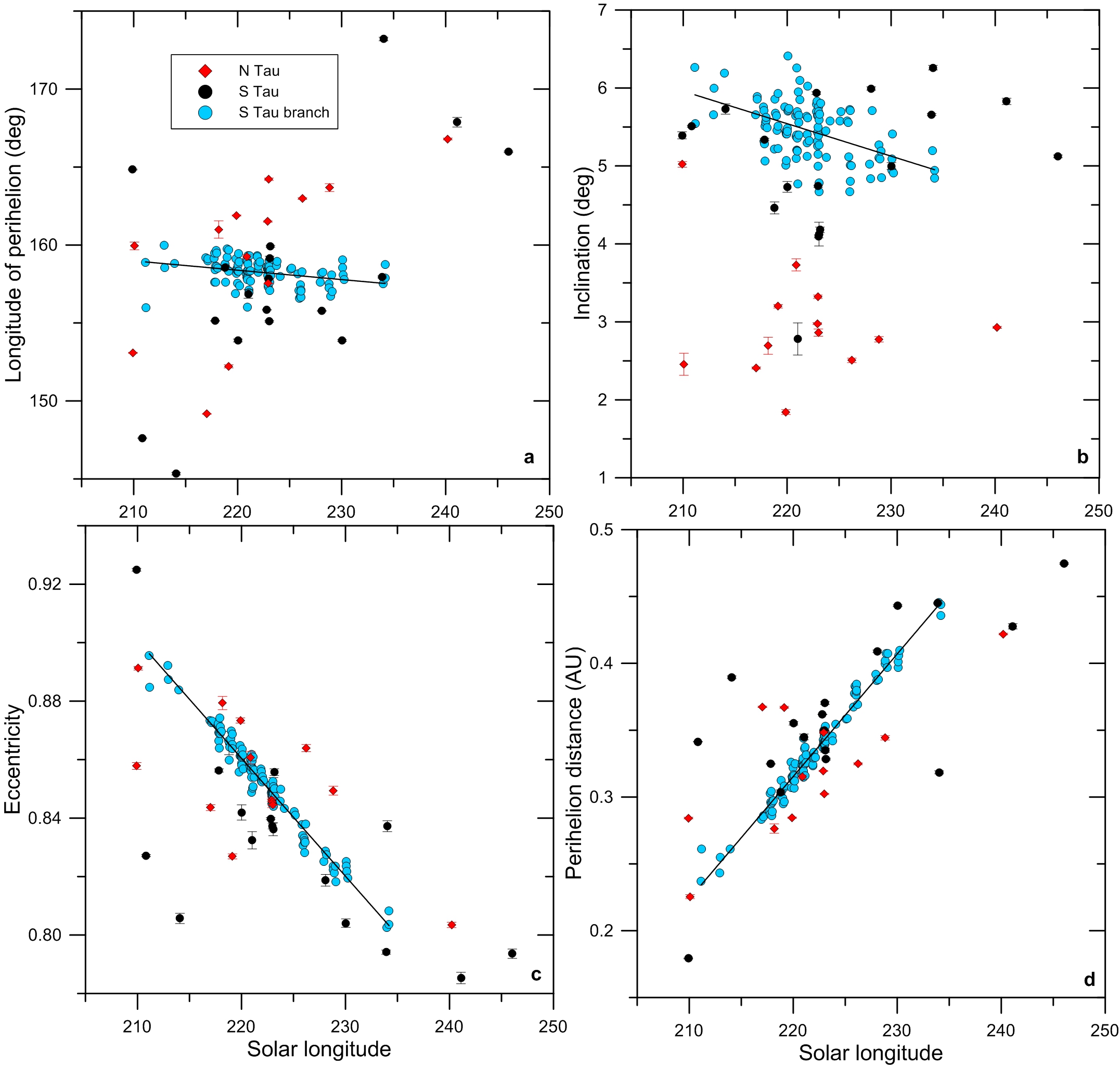}
\caption{Selected orbital elements as a function of solar longitude for Taurids observed in 2015. The symbols are the same as in
Fig.~\protect\ref{radiant}. Errors are in most cases smaller than symbol sizes and for clarity are plotted only for regular Taurids.}
\label{elements}
\end{figure*}

Longitude of perihelion, inclination, eccentricity, and perihelion distance as a function of solar longitude are plotted in Fig.~\ref{elements}.
Most notably, there is a concentration of orbits with longitude of perihelion, $\pi$ ($\pi=\Omega+\omega$, where $\Omega$ is longitude
of ascending node and $\omega$ is argument of perihelion), at $158\degr\pm2\degr$ (Fig.~\ref{elements}a).
There is only a weak correlation with solar
longitude. Similarly, there is a concentration of orbits with inclinations of $5.5\degr\pm1\degr$ (Fig.~\ref{elements}b).
Regular Taurids show much larger spread, 145 -- 175$\degr$ in $\pi$ and 2 -- 7$\degr$ in inclination.

Eccentricities and perihelion distances of the members of the new branch are steep functions
of solar longitude (Figs.~\ref{elements}c,d), i.e., 
\begin{eqnarray}
e & = & 0.901 - 0.00403\cdot (\lambda_{\sun} - 210\degr) \label{efunc} \\
q & = & 0.224 + 0.0092\cdot (\lambda_{\sun} - 210\degr). \label{qfunc}
\end{eqnarray}
All eccentricities lie within 0.012 from ~(\ref{efunc}) and perihelia lie within 0.027 AU from~(\ref{qfunc}).
Again, regular Taurids show much larger scatter.

\begin{figure}
\centering
\includegraphics[width=0.9\hsize]{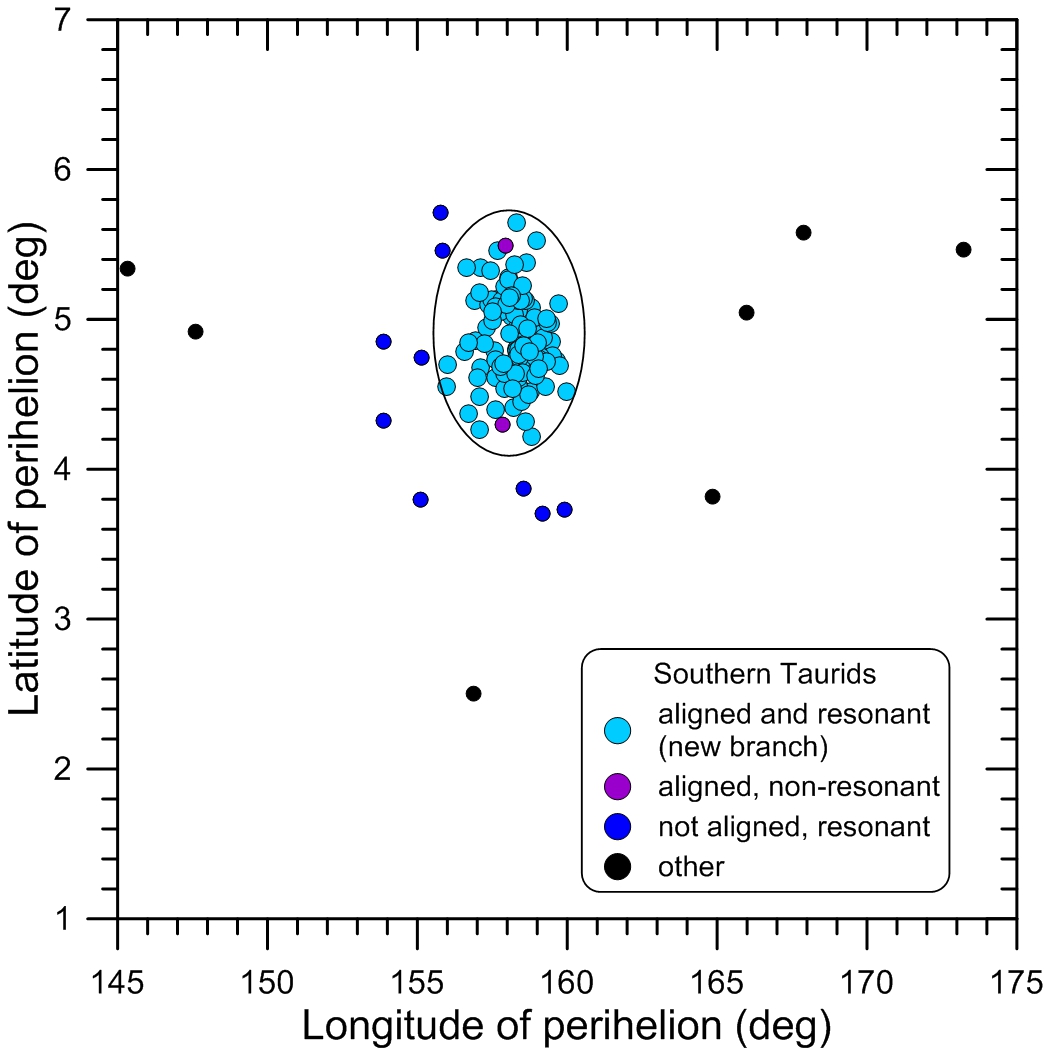}
\caption{Orientation of perihelia (latitude versus longitude) for Southern Taurids observed in 2015. This plot was used to define
the limits of the new branch, as indicated by the ellipse. The fireballs that fell within these limits but had
different semimajor axes (outside the 7:2 resonance) are plotted in purple. The fireballs outside these limits but
within the resonance are plotted in dark blue.}
\label{pi-beta}
\end{figure}

The new branch is best recognized in the plot of longitude of perihelion, $\pi$, versus latitude of perihelion,
$\beta$ ($\sin\beta=\sin\omega \sin i$, where $i$ is inclination), presented in Fig.~\ref{pi-beta}.
We can state that the new branch has $\pi$ between 155.9 -- 160$\degr$ and $\beta$ between 4.2 -- 5.7$\degr$.
For regular Southern Taurids the observed spread in $\beta$ is 2.5 -- $6.5\degr$.
Northern Taurids have negative $\beta$.

\begin{figure}
\centering
\includegraphics[width=0.9\hsize]{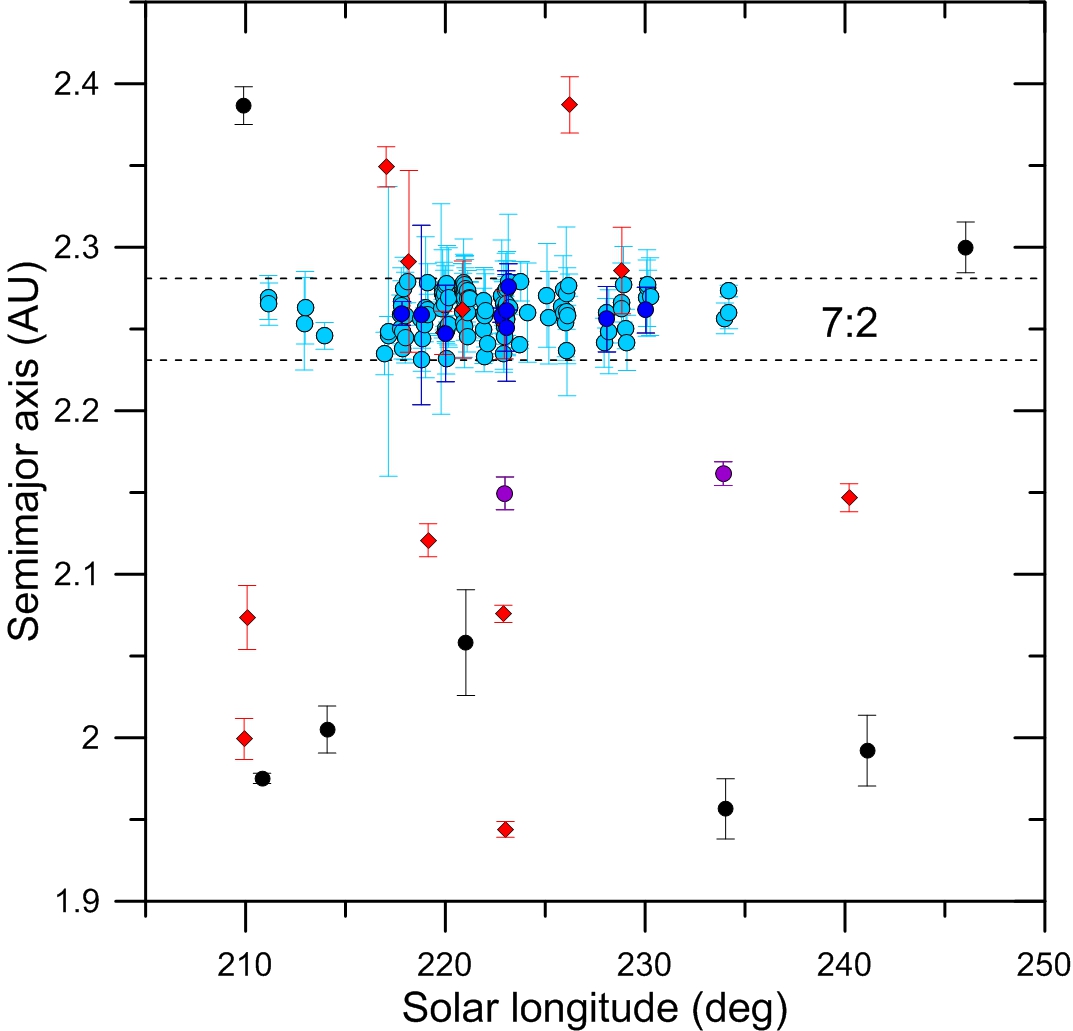}
\caption{Semimajor axis as a function of solar longitude for Taurids observed in 2015. For symbol explanation see
Figs.~\protect\ref{radiant} and \protect\ref{pi-beta}. Error bars are plotted for all fireballs.
The extent of the 7:2 resonance with Jupiter according to \citet{QJRAS} is indicated.}
\label{aaxis}
\end{figure}

Semimajor axes are plotted in Fig.~\ref{aaxis}. For regular Taurids, they lie between 1.9 and 2.4 AU.
According to the model of \citet{QJRAS}, the enhanced activity is caused by meteoroids trapped
in the 7:2 resonance with Jupiter. The resonance is located at 2.256 AU and extends from about 2.231 AU to 2.281 AU \citep{QJRAS}.
With two exceptions, the semimajor axes of all meteoroids with longitudes and latitudes of perihelia within the above-defined limits
fall in the 7:2 resonance. Only two meteoroids had significantly lower semimajor axes, 2.15 -- 2.16 AU. 
We consider them to be interlopers from the background population of Southern Taurids, although the 15:4 resonance located
at 2.155 AU might be at work here.

On the contrary, some Southern Taurids with perihelia outside the new branch limits were also in the 7:2 resonance.
As seen in Fig.~\ref{pi-beta}, all of the Souther Taurids had an orientation of perihelia relatively close to the new branch. 
Nevertheless, some Northern Taurids were in the 7:2 resonance as well and they were far from the new branch.

There is no correlation between semimajor axis and solar longitude.  The Tisserand parameter with respect to Jupiter increases with solar longitude from 2.9 to 3.1 within the new branch. This is due to the decreasing eccentricity. The often cited boundary
at $T_{\rm Jup}=3$  or 3.05 \citep[e.g.,][]{Tancredi} has no significance in this case.

According to the above definitions based on perihelion orientation and semimajor axis, there are 13 Northern Taurids in our data set,
18 regular Southern Taurids, and 113 members of the new branch.

\begin{figure}
\centering
\includegraphics[width=0.9\hsize]{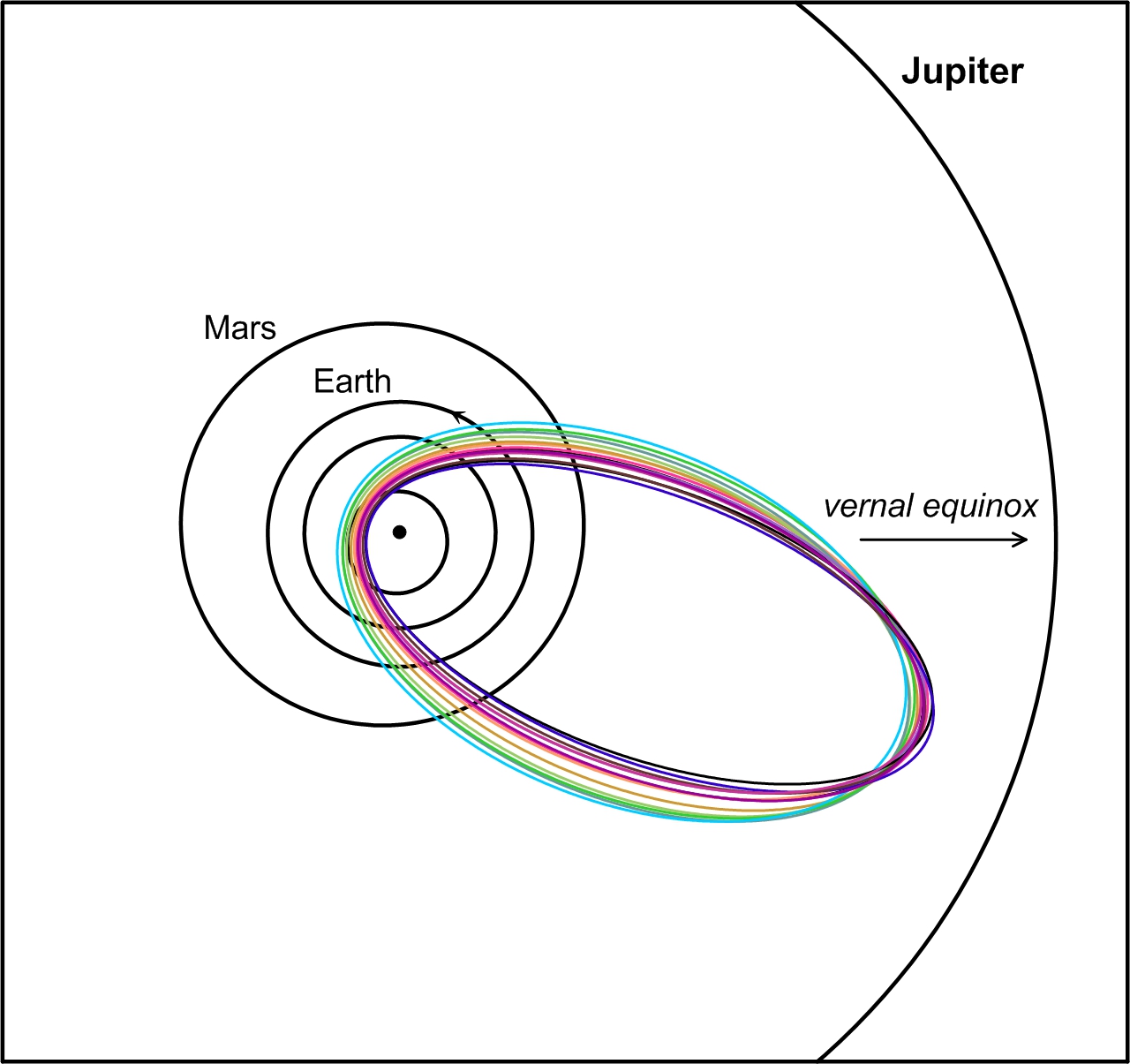}
\caption{Selected orbits of the Taurids from the new branch projected to the plane of ecliptic.}
\label{orbits}
\end{figure}

It is evident that the new branch represents an orbital structure that is much more compact than regular Taurids.
Since the activity of the new branch lasted almost one month, it cannot, however, be a narrow filament. In order to
visualize the new branch, we plotted  selected orbits covering the whole
activity period in Fig.~\ref{orbits}. Unlike usual meteoroid streams, where the orbits near perihelion largely overlap, here we see a concentric ring of orbits near perihelion, which is more than 0.2 AU wide. As the Earth moves around the Sun, it encounters first the orbits
with smaller perihelia and larger eccentricities. With increasing solar longitude, orbits with progressively larger perihelia and smaller
eccentricities are encountered. Since all of the semimajor axes are similar, eccentric orbits have larger
aphelia than less eccentric orbits and the orbits therefore intersect at about 3.6 AU.

\begin{figure}
\centering
\includegraphics[width=0.82\hsize]{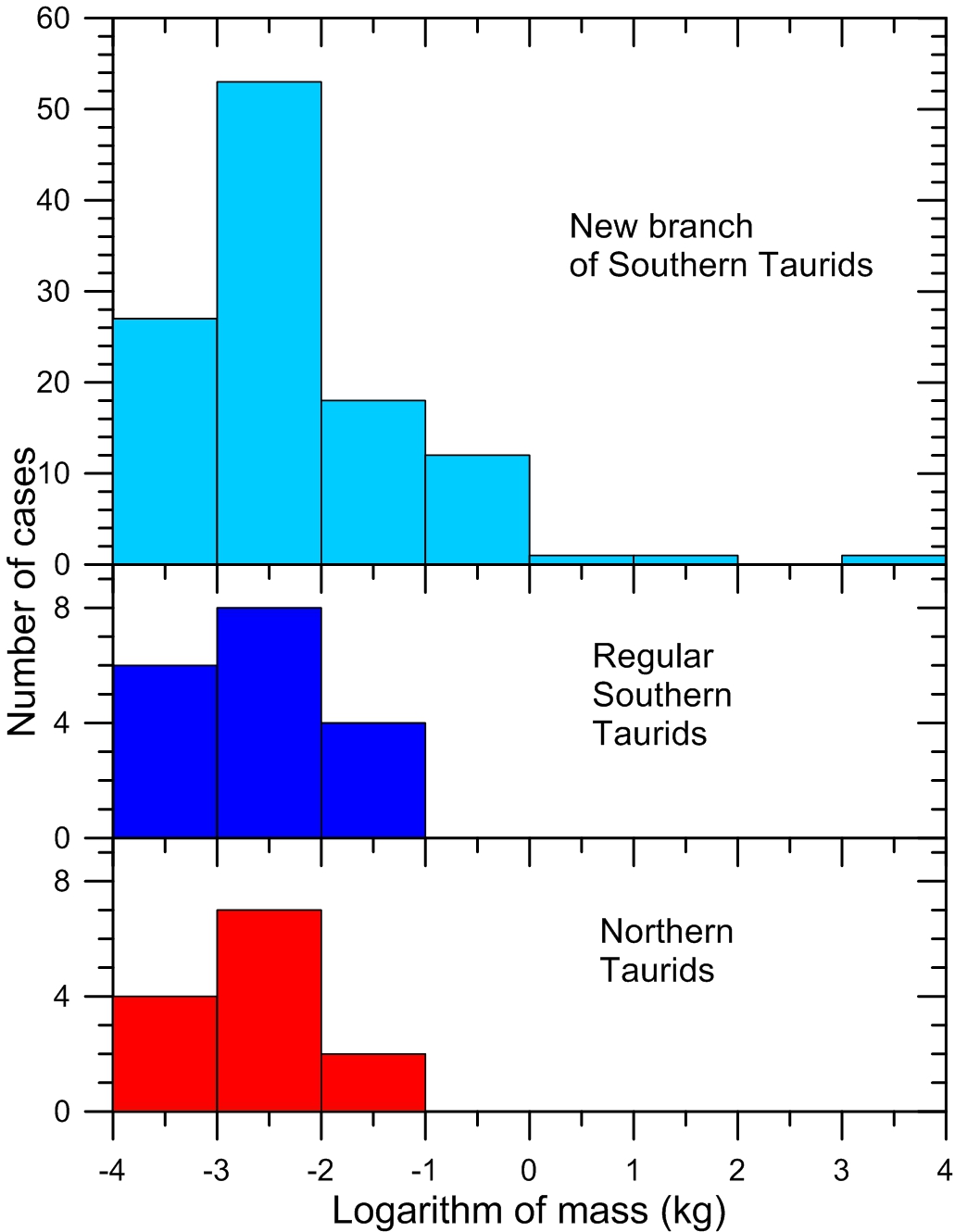}
\caption{Number of fireballs as a function of photometric mass for
Northern Taurids, regular Southern Taurids, and the new branch.}
\label{masses}
\end{figure}

\begin{figure}
\centering
\includegraphics[width=0.9\hsize]{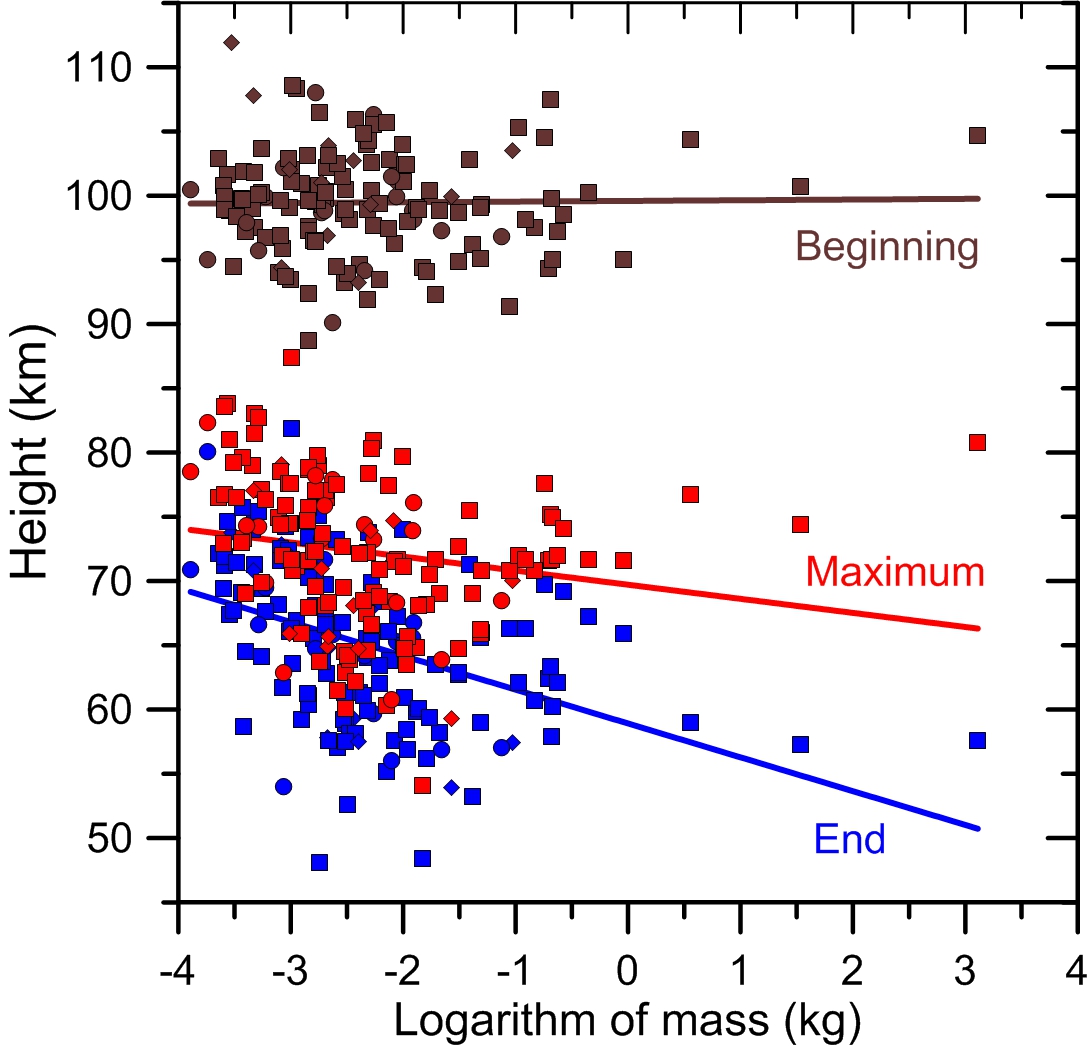}
\caption{Fireball heights at beginning, end, and maximum light as a function of photometric mass.
Northern Taurids are plotted as diamonds, regular Southern Taurids as circles, and new branch members as squares.}
\label{heights}
\end{figure}

\subsection{Physical properties}

The Taurids in our sample reached maximum absolute magnitude between $-2$ and $-18.6$. The photometric masses range from
0.1 gram to 1300 kg, i.e.,\ there is a range of 7 orders of magnitude in mass.
The mass distribution is given in Fig.~\ref{masses}, which shows that the new branch has a higher proportion of massive meteoroids.
The data in Fig.~\ref{masses} are biased because brighter meteors could be observed over large distances and under worse
conditions than faint meteors, nevertheless, the bias is the same for all branches.

\begin{figure}
\centering
\includegraphics[width=0.9\hsize]{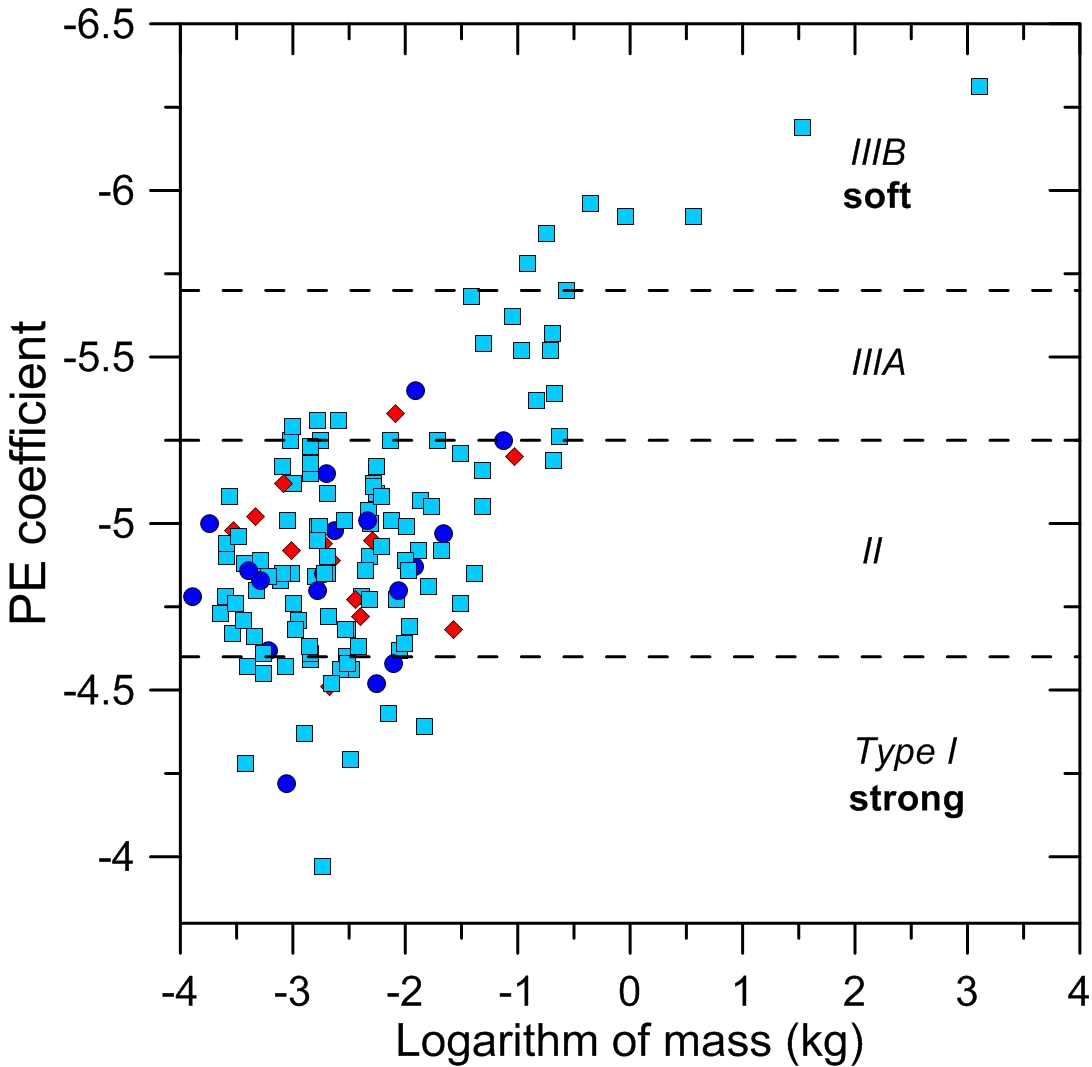}
\caption{Value of PE criterion \citep{PE} as a function of photometric mass for all observed Taurids. Northern Taurids are plotted as diamonds,
regular Souther Taurids as circles, and new branch members as squares. The dashed horizontal lines define the types I, II, IIIA, and IIIB.}
\label{PEgraf}
\end{figure}

The beginning, maximum brightness, and end heights of all studied fireballs
are plotted as a function of photometric mass in Fig.~\ref{heights}. These heights are good proxies to meteoroid structure,
although they depend to some extent
on observational circumstances (e.g.,\ range to the fireball) and on the slope of the trajectory.
Beginning heights show no dependence on mass and are generally between 90 and 110 km. For consistency we
use only data from digital all-sky cameras in the plot. The two brightest fireballs were captured by the narrow-field 
cameras at higher altitudes (see Table~\ref{twobolatm}). On the other hand, both these fireballs were located far from the all-sky cameras;
the beginning of EN\,311015\_180520 was 390 km from the closest camera and the beginning of EN\,311015\_231301  was 270 km distant.
If observed from closer distances, the beginnings would lie somewhat higher.

The maximum and end heights show large scatter. Many fireballs exhibited multiple flares of similar brightness.
Nevertheless, there were differences in physical properties of the meteoroids. This fact is mostly evident from the end heights.
There are differences of 25 km or more for meteoroids of similar masses. The expected trend of deeper penetration for larger
bodies is only weakly present. The lowest end heights (below 50 km) were achieved by two quite small meteoroids.
There are no obvious differences in physical properties between different branches of the stream.

\begin{figure}
\centering
\includegraphics[width=0.9\hsize]{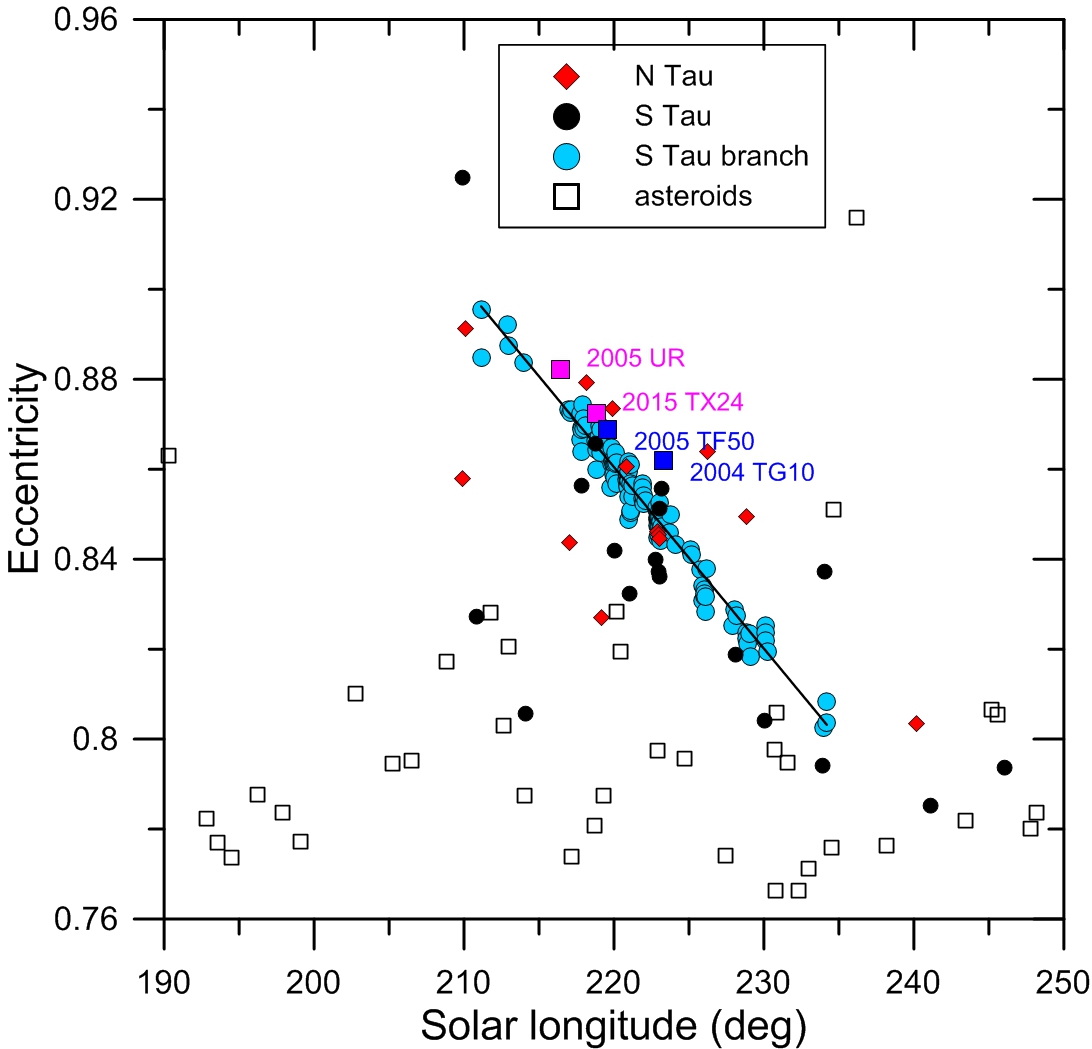}
\caption{Orbital eccentricity as a function for solar longitude at the closest approach to the Earth's orbit for 2015 Taurid fireballs and
asteroids from JPL database. Asteroids, which are likely related to the new Taurid branch are highlighted in magenta. Asteroids for which
the relation to the new branch was considered but not confirmed are shown as filled rectangles. They may be
related to other parts of the Taurid complex.}
\label{compar-e}
\end{figure}

Since the end height depends not only on the meteoroid properties but also on trajectory slope and entry speed, the PE criterion \citep{PE},
which compensates for these effects, can be used to better evaluate meteoroid strengths. According to the PE criterion, meteoroids
are classified into four types: I, II, IIIA, and IIIB \citep{Cep88}. Type I corresponds to stony meteorites and type IIIB to soft cometary
material. Figure~\ref{PEgraf} shows Taurid PE classification as a function of mass. We can see that Taurids cover all four types,
with a clear trend of larger meteoroids being more fragile. Most of meteoroids smaller than 30 grams belong to type II.
Some meteoroids with masses on the order of one gram clearly belong to type I. On the other hand, most  meteoroids
above 30 gram belong to type IIIA or IIIB and  only type IIIB is present above 300 gram. The fact that the two largest meteoroids
were very fragile was confirmed by fragmentation modeling (Sect.~\ref{twobrtau}).
Significant differences between small and large meteoroids suggest the existence of some
hierarchical structure and will be subject of future studies.

Similar heterogeneity of Taurid physical properties was observed recently by \citet{Matlovic}.
\citet{Brown} reported a Taurid that penetrated down to 35 km and \citet{Madiedo} and another one reaching
42.5 km. These authors suggested that Taurids might drop meteorites. Our data do not seem to support this possibility, since
at least a $\sim 1$ kg type I Taurid meteoroid would be needed to produce any meteorites. The SPMN\,051010 fireball
observed by \citet{Madiedo} on October 5, 2010 had a semimajor axis 3.0 AU and perihelion 0.47 AU. It may not be
Taurid at all. The SOMN 101031 fireball observed by \citet{Brown}, with a semimajor axis 2.9 AU, was also not a typical Taurid.

All atmospheric and physical data are given in Table~\ref{tbphysical}.

\begin{table*}
\caption{Orbital elements of asteroids discussed here as taken from the JPL database and converted to J2000.0 equinox.}
\label{tbasteroids}
\begin{tabular}{lllllllllr}
\hline
Asteroid & $\lambda_{\sun}$ & $a$ & $e$ & $q$ & $i$ & $\omega$ & $\Omega$ & $\pi$ & \multicolumn{1}{c}{$\beta$} \\ \hline \\[-3mm]
2005 UR        &   216.44&   2.254 &  0.882 &   0.266 &  6.94 &  140.40 & 20.03 & 160.43 & 4.42\\
2015 TX24     &   218.81 &  2.269  & 0.872  & 0.290  &  6.05  & 126.80 & 32.99 & 159.79 & 4.84\\
\hline \\[-3mm]
2005 TF50     &    219.60 &  2.272 &   0.869 &   0.298 &    10.70 &  159.67 & 0.66 & 160.33 & 3.70\\
2004 TG10     &  223.32   &2.234   & 0.862  & 0.308    &  4.18 & 317.11 & 205.13 & 162.23 & $-2.84$\\
\hline
\end{tabular}
\end{table*}

\begin{figure*}
\centering
\includegraphics[width=0.9\hsize]{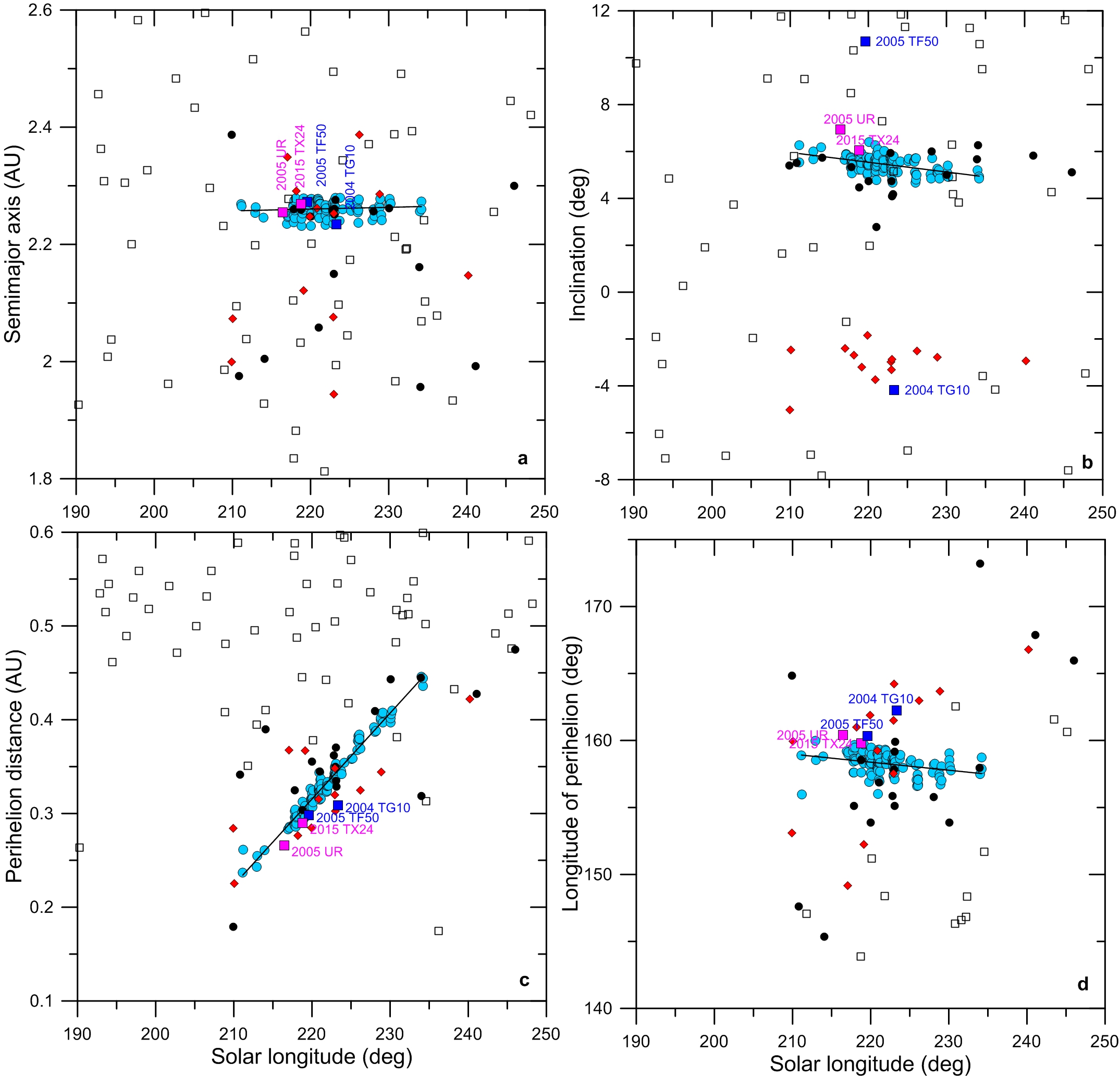}
\caption{Semimajor axis, inclination, perihelion distance, and longitude of perihelion
as a function for solar longitude at the closest approach to the Earth's orbit for 2015 Taurid fireballs and
asteroids from JPL database. Asteroids, which are likely related to the new Taurid branch are highlighted.
The symbols are the same as in Fig.~\protect\ref{compar-e}. The inclinations of Northern Taurids and asteroids
with $\Omega>180\degr$, which encounter the Earth near their descending node in October/November, are plotted as negative.
}
\label{comparall}
\end{figure*}

\section{Related asteroids}
\label{asteroids}

We performed a search for asteroids with orbits similar to the new Taurid branch responsible for the enhanced
activity in 2015. For that purpose, asteroids with $q<0.6$ AU, 1.8 AU $ < a < $ 2.8 AU, and $i<12\degr$ were selected from the
JPL Small-Body Database\footnote{http://ssd.jpl.nasa.gov/sbdb\_query.cgi, accessed January 25, 2017}. There are 329 such asteroids
known. We then plotted
selected orbital elements as a function of solar longitude at Earth Minimum Orbit Intersection Distance (MOID) to be compared with the observed fireballs. For fireballs we used solar longitude at the time of impact as the independent variable.
Since the asteroids
did not impact Earth and their orbits do not intersect Earth's orbit, we used for comparison the solar longitude, as
seen from the asteroid at the time when the asteroid is closest to the Earth's orbit.

Figure~\ref{compar-e} shows the comparison plot for eccentricity. We see that there is nearly random distribution of asteroids
with eccentricities smaller than 0.84 in the solar longitudes of interest. At higher eccentricities (0.86 -- 0.88), however, there is a noticeable
concentration of four asteroids (2005 UR, 2015 TX24, 2005 TF50, and 2004 TG10)
near solar longitude of $220\degr$. This concentration overlaps
with the new Taurid branch. Moreover, it follows the same trend of decreasing eccentricity with increasing solar longitude.

\begin{figure}
\centering
\includegraphics[width=0.9\hsize]{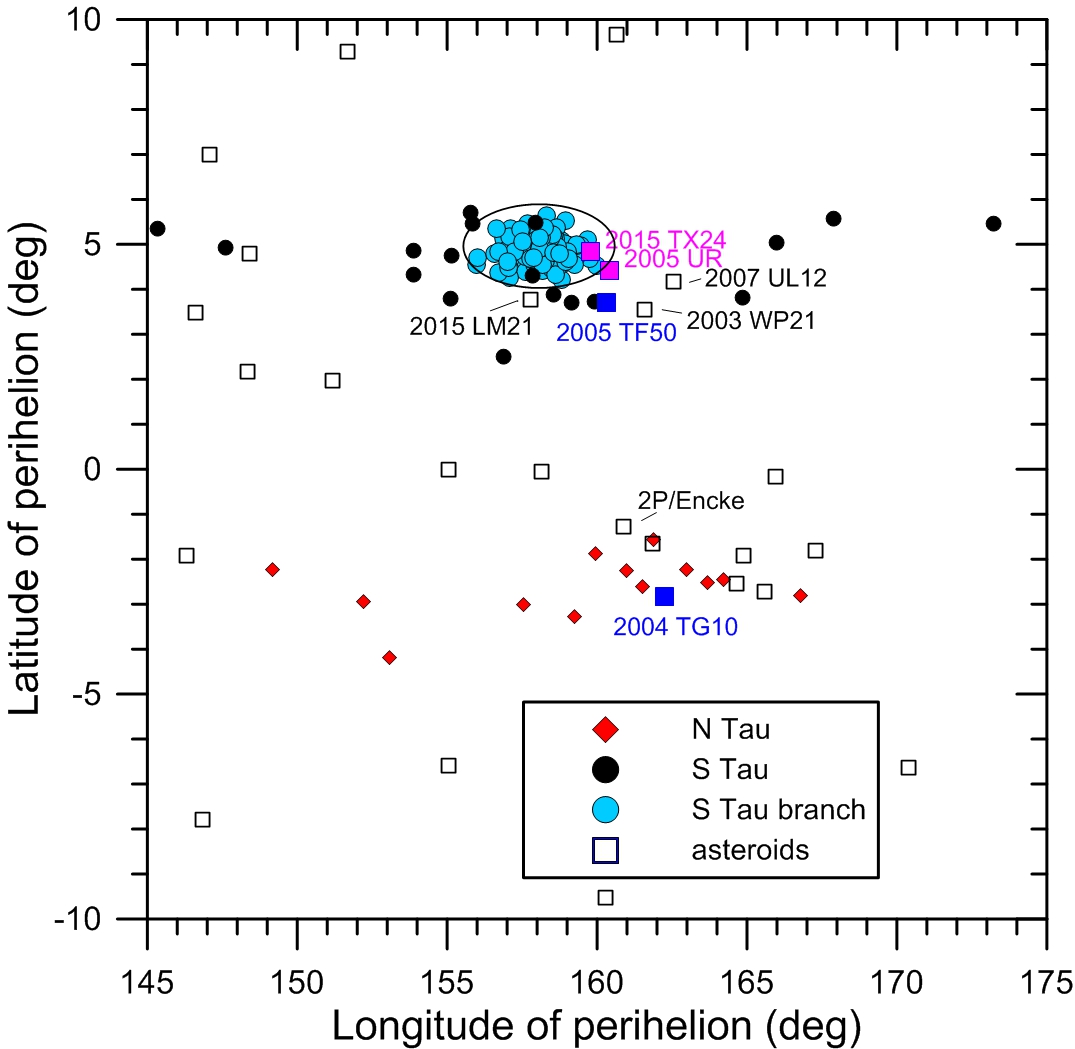}
\caption{Comparison of perihelia orientation of 2015 Taurids with asteroids and comets from the JPL database.}
\label{pi-beta-ast}
\end{figure}

Other orbital elements are compared in Fig.~\ref{comparall}. Perihelion distance is basically
a mirror image of eccentricity. Semimajor axes of all four asteroids of interest fall within the Taurid branch range, i.e.,\
also within the 7:2 resonance. As for inclination,
only 2015 TX24 falls exactly within the Taurid branch range. The 2005 UR asteroid is somewhat off but only about a half
degree from the edge of the Taurid branch. However, Taurid fireballs represent the part of the stream, which intersects
Earth's orbit. The whole stream is probably somewhat wider, so we consider it likely that 2005 UR is also part
of the stream. The 2005 TF50 asteroid matches all other elements very well but has an inclination of $10.7\degr$, i.e.,\
more than 4 degrees from the edge of the Taurid branch.
On the other hand, the orientation of perihelion is not so far from the new Taurid branch  (Fig.~\ref{pi-beta-ast}).
But the orbit of 2004 TG10 is oriented in the opposite way relative to the ecliptic. This asteroid may be in fact related to Northern Taurids.
At least two asteroids, 2015 TX24 and 2005 UR, are therefore good candidates for direct membership
in the new branch of Southern Taurids.

Asteroid 2015 TX24 was discovered by Pan-STARRS 1 on October 8, 2015 and was observed for 18 days in October 2015.
It passed closest to the Earth's orbit on October 28, 2015, i.e.,\ during the enhanced Taurid activity.
The MOID of the Earth is 0.010 AU.
The asteroid has an absolute magnitude of $H = 21.5$, which corresponds to diameter 200 -- 300 meters, assuming
albedo in the range 0.10 -- 0.05.

Asteroid 2005 UR was discovered by the Catalina Sky Survey on October 23, 2005 and
was observed for six days in October 2005. The MOID of the Earth is 0.034 AU. The absolute magnitude is $H = 21.6$, i.e.,\ very
similar to that of 2015 TX24. Asteroid 2005 UR approached the Earth's orbit at the end of December 2015 and was
therefore only 17$\degr$ in mean anomaly behind the Taurids observed in 2015. Moreover, as noted
by \citet{Olech}, Taurid activity was also enhanced when the asteroid passed close to the Earth in October 2005.

\begin{figure}
\centering
\includegraphics[width=0.9\hsize]{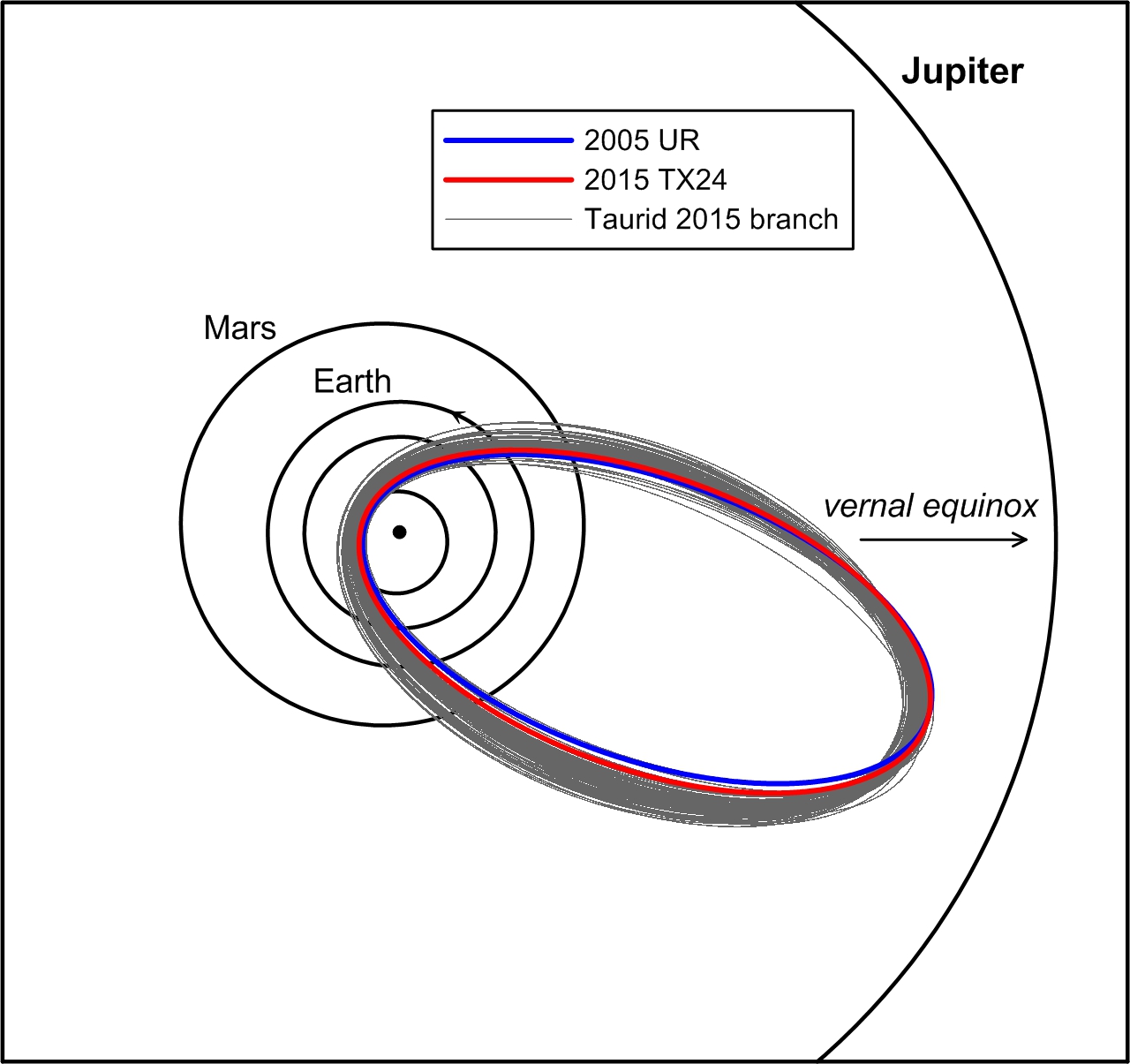}
\caption{Orbits of 2005 UR and 2015 TX24 in comparison with all Taurids orbits from the new branch (gray).}
\label{orbcompar}
\end{figure}

The orbits of both 2005 UR and 2015 TX24 are plotted in Fig.~\ref{orbcompar} together with the fireball orbits.
There is a good overlap. Orbital elements
of all four asteroids discussed here are given in Table~\ref{tbasteroids}.
Asteroid 2004 TG10 is a large object with $H = 19.4$. The albedo is very low and
the diameter was estimated to be $1.40 \pm 0.51$ km \citep{Nugent}. The possible relation of this asteroid to the Taurids was suggested already by
\citet{Jennbook}, \citet{Porub06}, and \citet{Babadz08}.  Asteroid 2005 TF50, with  $H = 20.3$, is of intermediate size. Its relation
to comet 2P/Encke and the Taurids was proposed by \citet{Porub06} and \citet{Olech}.

\section{Discussion}
\label{discussion}

We presented probably the most precise Taurid orbits obtained to date. Thanks to the sufficient precision of semimajor axes,
the theory of \citet{QJRAS} and \citet{Asher98} that the meteoroids responsible for enhanced Taurid activity
are in 7:2 resonance with Jupiter could be confirmed (at least for 2015 meteors).
This fact cannot be revealed from lower precision data such as those of \citet{Matlovic}\footnote{We observed 10 fireballs from their sample
and their semimajor axes are often off with respect to ours by several tenths of AU.}.
Moreover, we found that
the Taurid branch, which is responsible for the enhanced activity in 2015, forms an interesting orbital structure. Although the enhanced activity
lasted for 23 days according to our data, all orbits had very similar orientation of the line of apsides, i.e.,\ the
longitude and latitude of perihelion. Since semimajor axes were in a narrow range and
all observed meteoroids  had to intersect Earth's orbit, only one free parameter remains. That is why there is a good correlation between the
longitude of the ascending node (or, equivalently,  solar longitude at the date of observation), eccentricity, and perihelion distance.

There was, nevertheless, some spread of orbital elements within the new branch. The longitudes of perihelia were within the range
155.9 -- $160\degr$ and latitudes of perihelia within the range 4.2 -- $5.7\degr$. The semimajor axes were
within the resonance limits, 2.23 -- 2.28 AU. The additional
condition for the new branch membership follows from the limited period of activity and can be expressed, for example,\ in terms of eccentricity
lying between 0.80 -- 0.90. 
Three asteroids, 2015 TX24, 2005 UR, and 2005 TF50, fully or nearly satisfy all these conditions.
Fig.~\ref{pi-beta-ast} shows three other asteroids (2003 WP21, 2007 UL12, and 2015 LM21)
with perihelia orientation not far from the new branch, but none of them simultaneously 
fulfills both the semimajor axis and eccentricity criteria.

Since the Earth does not encounter the new branch every year, it is evident that meteoroids of the new branch are not spread along
the whole orbit. The model and observations of \citet{QJRAS} and \citet{Asher98} suggest that the enhanced activity of Taurids
is caused by a resonant swarm of meteoroids, which extends $\pm$ 30 -- 40$\degr$ from the center of the swarm in mean anomaly.
It does not, however, necessarily mean
that the new branch observed in 2015 is identical to or representative of the whole swarm.
The orbits of meteoroids observed by the EN during the enhanced
activity in 1995 had somewhat different characteristics than in 2015; these had larger semimajor axes
and smaller perihelia, which did not change so much with solar longitude.

The new branch contains quite large bodies. Our brightest fireball was caused by a body in excess of 1000 kg, which corresponds to
diameter more than one meter, assuming that bulk density was not higher than 2000 kg m$^{-3}$. This body was disintegrated very high in the atmosphere and likely had high porosity and low bulk density. The
NASA JPL fireball page\footnote{http://neo.jpl.nasa.gov/fireballs/, accessed February 3, 2017} lists a fireball with 10 times higher
radiated energy, which  occurred on the same day (October 31, 2015 11:34:30 UT) above the Pacific Ocean at a quite large height of 71 km.
Considering the unusual height, it is likely that that fireball belonged to the Taurid
new branch as well. The size of that body was 2-3 meters or more. Two similar, slightly smaller, events occurred on November 2, 2005
(05:16:47 and 07:04:32), also over Pacific Ocean. The heights of these bolides were 74 and 68.5 km, respectively. These three fireballs
are among the top five events with largest heights among the 288 fireballs with known heights listed at the NASA JPL page. Their
trajectories and velocities are not given but the Taurid radiant was above the horizon in all cases. We note that 2005 was also a year of
enhanced Taurid activity.

Fireball data therefore prove the presence of meter-sized bodies among the Taurid new branch. Based on orbital similarity,
we argue that asteroids of several hundred meters in diameter are members of the Taurid new branch as well. This is almost certain
for 2015 TX24, very likely for 2005 UR, and possible for 2005 TF50. We are not speaking about a distant relationship.
The discovered Taurid branch is simply a population of bodies with the size range from several millimeters to several hundred of
meters, which all move together around the Sun. Every few years, the Earth is encountering this branch for a period of about
three weeks. During that time, the chance of impact of an asteroid of significant size (tens of meters) is significantly enhanced.
Even if intrinsically weak, bodies of such size can penetrate deep in the atmosphere \citep{Shuvalov} and pose a hazard to the ground.

We will allow theoretical celestial mechanicians to explain the formation and evolution of the Taurid new branch and the Taurid complex
as a whole.
A structure similar to the new branch could be created by a disruption of a parent body at heliocentric distance of about 3.6 AU  (where
the orbits come close together) but ejection velocities up to 1.5 km s$^{-1}$ and subsequent removal of all non-resonant orbits
would be needed.
Also, asteroids 2005 TF50,  2015 TX24,  and 2005 UR  can all be related to 2004 TG10  but located at a different
phase along the secular cycle as computed for 2004 TG10 by \citet{Porub06}. Asteroid 2005 TF50 is about 2000 years behind,
2005 UR is about 2300 years behind, and 2015 TX24  is about 2400 years behind 2004 TG10. The elements $\omega$, $\Omega$,
and $i$ all agree well with this assumption. For 2005 TF50 and 2015 TX24 $e$ and $q$ are also in agreement.
The new Taurid branch can be also part of this relation. In fact, the orbital elements of the theoretical
Southern Taurid meteors derived from 2004 TG10,  as computed by \citet{Babadz08}, fall perfectly among the
Taurid branch fireballs in Fig.~\ref{elements}. Only in $\pi$ there is a difference of $2.5\degr$. But only the central
part of the new branch at $\lambda_{\sun} \sim 220\degr$ can be explained in this way.

\section{Conclusions}

We presented data of unprecedented precision for a large sample of 144 Taurid fireballs observed by the European fireball
network in 2015. This data set contains precise and detailed data on the Taurids covering 7 orders in mass, i.e., from tenths of a gram to one-ton meteoroids. We have shown that the enhanced Taurid activity in 2015 was produced by a well-defined branch embedded within the much broader
Southern Taurid stream. The new branch can be characterized by the longitudes of perihelia lying between 155.9 -- 160$\degr$,
latitudes of perihelia between 4.2 -- 5.7$\degr$, semimajor axes between 2.23 -- 2.28 AU, and eccentricities between 0.80 -- 0.90.
These orbits form a concentric ring in the inner solar system with perihelia between 0.23 -- 0.45 AU. The new branch lies within the semimajor axis range spanned by the 7:2 resonance, indicating strongly that the meteoroids responsible for the outbursts are within this resonance, as expected from the model of \citet{QJRAS}.
The Earth was the encountering members of the new branch at their ascending nodes between October 25 and November 17.
The orbital configuration of the branch cause meteoroids with progressively lower eccentricities, 
larger perihelion distances, and lower entry
velocities to encounter the Earth during the activity period.

The explanation of the structure and evolution of the new branch and its relation to the whole Taurid complex must be
left to future theoretical studies. Nevertheless, we confirm earlier observations that the Taurid stream contains large
meteoroids. This is valid for the new branch in particular. The largest object we observed was at least one meter in diameter.
A ten times more massive object observed on the same day over the Pacific Ocean probably belonged to this new branch
as well. Moreover, the orbits of asteroids 2015 TX24 and 2005 UR, both of diameters of several hundreds of meters, place them within
the new Taurid branch as well. It is therefore very likely that the branch also contains numerous objects of decameter size.
Although our data show that large Taurids have porous and fragile structure, objects of tens or hundreds of meters in size pose a
hazard to the ground even if they have low intrinsic strength. Theoretical and observational studies and searches
for related asteroids belonging to this newly discovered and described branch of Southern Taurids are therefore highly recommended. A better understanding of this real source of potentially hazardous objects that are large enough to cause significant regional or even continental damage on the Earth is a task of capital importance.

\begin{acknowledgements}

This work was supported by the Praemium Academiae of the Czech Academy of Sciences, grant 16-00761S from the Czech Science Foundation,
and the Czech institutional project RVO:67985815. Operation of the Slovak station Star\' a Lesn\' a was supported by the project ITMS No.
26220120029, based on the supporting operational Research and development program financed from the European Regional Development Fund. We thank especially J. Kecl\' ikov\' a, but also L. Shrben\' y and H. Zichov\' a for a careful measuring of all photographic images. We also thank B. Pelc, T. Chl\'{\i}bec, L. Sklen\' ar and D. \v S\v cerba for their images of two brightest Taurids.  

\end{acknowledgements}

\begin{longtab}
\begin{longtable}{ccrrrrrrrrrr}
\caption{Radiant and orbital data for 2015 Taurid fireballs.  The code of each fireball also contains the date (in ddmmyy format) and GMT time corresponding to beginning rounded to whole second (in hhmmss format)}
\label{tborbits}\\
\hline\hline \\[-2mm]
Code & Branch\footnotemark[1] & \multicolumn{1}{c}{$\lambda_{\sun}$} & \multicolumn{1}{c}{$\alpha_{\rm g}$} &
 \multicolumn{1}{c}{$\delta_{\rm g}$} & \multicolumn{1}{c}{$v_{\rm g}$} &\multicolumn {1}{c}{$a$} & \multicolumn{1}{c}{$e$} &
 \multicolumn{1}{c}{$q$} & \multicolumn{1}{c}{$\omega$} &  \multicolumn{1}{c}{$i$} & \multicolumn{1}{c}{$\pi$}  \\
\hline \\[-2mm]
\endfirsthead
\caption{continued.}\\
\hline\hline \\[-3mm]
Code & Branch\footnotemark[1] & \multicolumn {1}{c}{$\lambda_{\sun}$} & \multicolumn {1}{c}{$\alpha_{\rm g}$} & \multicolumn {1}{c}{$\delta_{\rm g}$} & \multicolumn {1}{c}{$v_{\rm g}$} & \multicolumn {1}{c}{$a$} & \multicolumn {1}{c}{$e$} & \multicolumn {1}{c}{$q$} & \multicolumn {1}{c}{$\omega$} & \multicolumn {1}{c}{$i$} & \multicolumn {1}{c}{$\pi$}  \\
\hline \\[-2mm]
\endhead
\hline
\endfoot
EN231015\_204348&S & 209.908&  49.00&  15.31&  35.04&   2.387&  0.9249&  0.1792& 134.92&   5.39& 164.84\\[-0.3ex]
&&&                            0.01&   0.02&   0.03&   0.012&  0.0005&  0.0003&   0.03&   0.04&   0.03\\
EN231015\_211327&N & 209.928&  42.26&  20.04&  30.16&   1.999&  0.8578&  0.2842& 303.17&   5.02& 153.09\\[-0.3ex]
&&&                            0.02&   0.02&   0.06&   0.013&  0.0012&  0.0006&   0.03&   0.04&   0.03\\
EN241015\_004546&N & 210.075&  45.88&  18.83&  32.44&   2.074&  0.8913&  0.2255& 309.90&   2.46& 159.94\\[-0.3ex]
&&&                            0.12&   0.08&   0.03&   0.020&  0.0006&  0.0014&   0.24&   0.14&   0.24\\
EN241015\_185031&S & 210.825&  42.91&  11.61&  28.19&   1.975&  0.8272&  0.3414& 116.78&   5.51& 147.62\\[-0.3ex]
&&&                            0.01&   0.01&   0.01&   0.003&  0.0003&  0.0002&   0.02&   0.01&   0.02\\
EN251015\_022301&SB& 211.138&  47.62&  13.76&  32.63&   2.269&  0.8955&  0.2370& 127.73&   6.26& 158.88\\[-0.3ex]
&&&                            0.01&   0.04&   0.05&   0.013&  0.0008&  0.0005&   0.04&   0.06&   0.04\\
EN251015\_031725&SB& 211.176&  46.32&  13.59&  31.73&   2.265&  0.8847&  0.2612& 124.80&   5.54& 155.99\\[-0.3ex]
&&&                            0.03&   0.02&   0.05&   0.013&  0.0008&  0.0006&   0.06&   0.03&   0.06\\
EN261015\_213736&SB& 212.933&  48.99&  14.44&  32.37&   2.253&  0.8921&  0.2430& 127.03&   5.66& 159.97\\[-0.3ex]
&&&                            0.03&   0.03&   0.10&   0.028&  0.0017&  0.0010&   0.07&   0.05&   0.07\\
EN261015\_224031&SB& 212.977&  48.51&  13.96&  31.98&   2.263&  0.8874&  0.2549& 125.55&   5.99& 158.54\\[-0.3ex]
&&&                            0.03&   0.03&   0.08&   0.022&  0.0014&  0.0008&   0.06&   0.05&   0.06\\
EN271015\_220749&SB& 213.951&  49.26&  13.94&  31.74&   2.246&  0.8838&  0.2610& 124.85&   6.19& 158.82\\[-0.3ex]
&&&                            0.01&   0.05&   0.03&   0.008&  0.0005&  0.0004&   0.04&   0.07&   0.04\\
EN281015\_011855&S & 214.084&  43.91&  11.12&  26.78&   2.005&  0.8057&  0.3896& 111.24&   5.73& 145.34\\[-0.3ex]
&&&                            0.03&   0.07&   0.07&   0.014&  0.0018&  0.0010&   0.08&   0.07&   0.08\\
EN301015\_222401&SB& 216.958&  51.08&  14.52&  30.96&   2.235&  0.8733&  0.2832& 122.22&   5.65& 159.19\\[-0.3ex]
&&&                            0.02&   0.06&   0.05&   0.013&  0.0009&  0.0006&   0.05&   0.08&   0.05\\
EN311015\_002325&N & 217.040&  44.77&  19.16&  28.36&   2.349&  0.8436&  0.3674& 292.16&   2.41& 149.17\\[-0.3ex]
&&&                            0.01&   0.01&   0.04&   0.012&  0.0010&  0.0005&   0.02&   0.01&   0.02\\
EN311015\_023900&SB& 217.134&  51.15&  14.34&  30.90&   2.246&  0.8726&  0.2861& 121.83&   5.89& 158.98\\[-0.3ex]
&&&                            0.01&   0.03&   0.04&   0.011&  0.0008&  0.0005&   0.04&   0.04&   0.04\\
EN311015\_025717&SB& 217.147&  51.20&  14.38&  30.94&   2.249&  0.8732&  0.2850& 121.95&   5.86& 159.11\\[-0.3ex]
&&&                            0.52&   0.04&   0.03&   0.089&  0.0024&  0.0061&   1.03&   0.24&   1.03\\
EN311015\_172431&SB& 217.749&  51.56&  14.50&  30.82&   2.259&  0.8719&  0.2893& 121.40&   5.76& 159.17\\[-0.3ex]
&&&                            0.02&   0.03&   0.04&   0.012&  0.0008&  0.0004&   0.04&   0.05&   0.04\\
EN311015\_180520&SB& 217.777&  51.69&  14.59&  30.87&   2.250&  0.8724&  0.2872& 121.69&   5.71& 159.48\\[-0.3ex]
&&&                            0.01&   0.02&   0.03&   0.009&  0.0006&  0.0003&   0.02&   0.02&   0.02\\
EN311015\_182902&SB& 217.794&  50.83&  14.50&  30.37&   2.267&  0.8665&  0.3026& 119.81&   5.31& 157.62\\[-0.3ex]
&&&                            0.02&   0.03&   0.07&   0.021&  0.0015&  0.0007&   0.03&   0.03&   0.03\\
EN311015\_185530&SB& 217.812&  51.17&  14.68&  30.59&   2.265&  0.8693&  0.2960& 120.59&   5.28& 158.42\\[-0.3ex]
&&&                            0.01&   0.03&   0.05&   0.015&  0.0011&  0.0005&   0.03&   0.04&   0.03\\
EN311015\_192126&S & 217.830&  49.84&  13.99&  29.62&   2.260&  0.8563&  0.3247& 117.29&   5.34& 155.13\\[-0.3ex]
&&&                            0.01&   0.01&   0.03&   0.007&  0.0006&  0.0003&   0.02&   0.02&   0.02\\
EN311015\_200534&SB& 217.861&  51.36&  14.57&  30.59&   2.249&  0.8689&  0.2949& 120.78&   5.51& 158.65\\[-0.3ex]
&&&                            0.02&   0.07&   0.06&   0.017&  0.0012&  0.0007&   0.06&   0.09&   0.06\\
EN311015\_202117&SB& 217.872&  50.82&  14.66&  30.23&   2.238&  0.8640&  0.3044& 119.71&   5.06& 157.60\\[-0.3ex]
&&&                            0.01&   0.01&   0.00&   0.002&  0.0001&  0.0001&   0.02&   0.01&   0.02\\
EN311015\_211904&SB& 217.912&  51.80&  14.74&  30.96&   2.275&  0.8743&  0.2860& 121.73&   5.56& 159.65\\[-0.3ex]
&&&                            0.06&   0.02&   0.06&   0.019&  0.0011&  0.0009&   0.13&   0.04&   0.13\\
EN311015\_230919&SB& 217.988&  51.80&  14.69&  30.79&   2.245&  0.8713&  0.2889& 121.49&   5.58& 159.50\\[-0.3ex]
&&&                            0.03&   0.02&   0.06&   0.015&  0.0011&  0.0006&   0.06&   0.03&   0.06\\
EN311015\_231301&SB& 217.991&  51.44&  14.49&  30.59&   2.258&  0.8689&  0.2960& 120.62&   5.62& 158.62\\[-0.3ex]
&&&                            0.02&   0.04&   0.10&   0.027&  0.0020&  0.0010&   0.06&   0.06&   0.06\\
EN011115\_013625&SB& 218.091&  51.42&  14.58&  30.60&   2.279&  0.8696&  0.2971& 120.41&   5.49& 158.51\\[-0.3ex]
&&&                            0.01&   0.03&   0.05&   0.013&  0.0009&  0.0005&   0.03&   0.03&   0.03\\
EN011115\_033911&N & 218.176&  50.88&  20.51&  31.26&   2.291&  0.8793&  0.2765& 302.85&   2.69& 160.99\\[-0.3ex]
&&&                            0.28&   0.06&   0.11&   0.056&  0.0022&  0.0035&   0.56&   0.11&   0.56\\
EN011115\_174410&SB& 218.763&  51.87&  14.84&  30.40&   2.256&  0.8665&  0.3013& 120.00&   5.22& 158.77\\[-0.3ex]
&&&                            0.01&   0.02&   0.01&   0.004&  0.0003&  0.0002&   0.02&   0.02&   0.02\\
EN011115\_183646&S & 218.799&  51.61&  15.34&  30.31&   2.259&  0.8657&  0.3034& 119.74&   4.46& 158.55\\[-0.3ex]
&&&                            0.06&   0.06&   0.19&   0.055&  0.0040&  0.0019&   0.13&   0.08&   0.13\\
EN011115\_191104&SB& 218.823&  51.60&  14.15&  29.98&   2.231&  0.8599&  0.3127& 118.77&   5.86& 157.61\\[-0.3ex]
&&&                            0.01&   0.01&   0.03&   0.007&  0.0005&  0.0003&   0.02&   0.01&   0.02\\
EN011115\_200918&SB& 218.864&  51.96&  14.48&  30.28&   2.244&  0.8643&  0.3044& 119.68&   5.68& 158.56\\[-0.3ex]
&&&                            0.03&   0.06&   0.04&   0.013&  0.0009&  0.0006&   0.07&   0.08&   0.07\\
EN011115\_223909&SB& 218.968&  52.28&  14.64&  30.47&   2.253&  0.8671&  0.2994& 120.23&   5.64& 159.21\\[-0.3ex]
&&&                            0.01&   0.02&   0.04&   0.011&  0.0008&  0.0004&   0.03&   0.03&   0.03\\
EN011115\_234207&SB& 219.011&  52.47&  14.88&  30.64&   2.263&  0.8697&  0.2948& 120.72&   5.46& 159.75\\[-0.3ex]
&&&                            0.03&   0.03&   0.16&   0.043&  0.0032&  0.0016&   0.07&   0.05&   0.07\\
EN021115\_020950&SB& 219.114&  51.90&  14.79&  30.26&   2.278&  0.8652&  0.3071& 119.23&   5.22& 158.36\\[-0.3ex]
&&&                            0.02&   0.02&   0.04&   0.012&  0.0009&  0.0005&   0.04&   0.02&   0.04\\
EN021115\_021740&SB& 219.119&  51.96&  14.57&  30.19&   2.262&  0.8638&  0.3081& 119.18&   5.51& 158.31\\[-0.3ex]
&&&                            0.02&   0.05&   0.05&   0.014&  0.0011&  0.0007&   0.05&   0.06&   0.05\\
EN021115\_022525&SB& 219.125&  52.64&  14.55&  30.60&   2.259&  0.8688&  0.2964& 120.56&   5.93& 159.70\\[-0.3ex]
&&&                            0.03&   0.04&   0.05&   0.013&  0.0009&  0.0006&   0.06&   0.05&   0.06\\
EN021115\_024553&N & 219.139&  47.24&  20.59&  27.81&   2.121&  0.8269&  0.3671& 293.11&   3.20& 152.22\\[-0.3ex]
&&&                            0.05&   0.02&   0.04&   0.010&  0.0009&  0.0007&   0.10&   0.02&   0.10\\
EN021115\_182450&SB& 219.792&  51.84&  14.14&  29.61&   2.262&  0.8558&  0.3262& 117.09&   5.76& 156.90\\[-0.3ex]
&&&                            0.39&   0.19&   0.04&   0.064&  0.0022&  0.0046&   0.77&   0.28&   0.77\\
EN021115\_195540&SB& 219.855&  52.70&  14.82&  30.26&   2.274&  0.8649&  0.3072& 119.23&   5.44& 159.10\\[-0.3ex]
&&&                            0.02&   0.03&   0.02&   0.005&  0.0003&  0.0003&   0.04&   0.04&   0.04\\
EN021115\_201534&SB& 219.868&  52.22&  14.95&  29.97&   2.272&  0.8612&  0.3153& 118.31&   5.01& 158.19\\[-0.3ex]
&&&                            0.03&   0.03&   0.09&   0.027&  0.0020&  0.0010&   0.06&   0.04&   0.06\\
EN021115\_205431&N & 219.895&  52.44&  20.31&  30.88&   2.248&  0.8734&  0.2845& 302.04&   1.84& 161.90\\[-0.3ex]
&&&                            0.02&   0.02&   0.05&   0.013&  0.0009&  0.0005&   0.03&   0.03&   0.03\\
EN021115\_213614&SB& 219.925&  52.39&  14.96&  30.03&   2.270&  0.8620&  0.3134& 118.53&   5.07& 158.47\\[-0.3ex]
&&&                            0.02&   0.05&   0.07&   0.021&  0.0015&  0.0008&   0.05&   0.06&   0.05\\
EN021115\_215818&SB& 219.940&  52.49&  14.71&  30.05&   2.272&  0.8620&  0.3134& 118.52&   5.43& 158.48\\[-0.3ex]
&&&                            0.02&   0.07&   0.05&   0.014&  0.0010&  0.0006&   0.06&   0.09&   0.06\\
EN021115\_220435&SB& 219.944&  52.45&  14.70&  29.99&   2.265&  0.8611&  0.3146& 118.41&   5.41& 158.37\\[-0.3ex]
&&&                            0.01&   0.02&   0.05&   0.013&  0.0010&  0.0005&   0.03&   0.02&   0.03\\
EN021115\_232112&SB& 219.998&  52.14&  14.55&  29.78&   2.277&  0.8587&  0.3217& 117.55&   5.41& 157.56\\[-0.3ex]
&&&                            0.01&   0.02&   0.04&   0.012&  0.0009&  0.0005&   0.03&   0.03&   0.03\\
EN021115\_234348&S & 220.013&  50.43&  14.32&  28.59&   2.247&  0.8419&  0.3552& 113.84&   4.73& 153.87\\[-0.3ex]
&&&                            0.04&   0.06&   0.11&   0.029&  0.0026&  0.0013&   0.09&   0.07&   0.09\\
EN021115\_235259&SB& 220.020&  52.65&  14.71&  30.02&   2.250&  0.8610&  0.3128& 118.67&   5.48& 158.70\\[-0.3ex]
&&&                            0.01&   0.01&   0.03&   0.008&  0.0006&  0.0003&   0.02&   0.01&   0.02\\
EN031115\_002007&SB& 220.038&  52.51&  14.65&  29.83&   2.232&  0.8580&  0.3169& 118.27&   5.44& 158.33\\[-0.3ex]
&&&                            0.02&   0.03&   0.03&   0.009&  0.0007&  0.0004&   0.04&   0.04&   0.04\\
EN031115\_011247&SB& 220.075&  53.06&  14.70&  30.22&   2.249&  0.8636&  0.3068& 119.37&   5.70& 159.46\\[-0.3ex]
&&&                            0.02&   0.02&   0.06&   0.017&  0.0013&  0.0007&   0.04&   0.03&   0.04\\
EN031115\_012404&SB& 220.083&  52.76&  13.98&  30.02&   2.278&  0.8613&  0.3159& 118.21&   6.41& 158.31\\[-0.3ex]
&&&                            0.02&   0.05&   0.09&   0.023&  0.0018&  0.0010&   0.06&   0.07&   0.06\\
EN031115\_025102&SB& 220.143&  52.88&  14.59&  30.06&   2.253&  0.8616&  0.3119& 118.76&   5.72& 158.92\\[-0.3ex]
&&&                            0.02&   0.12&   0.04&   0.013&  0.0009&  0.0008&   0.10&   0.15&   0.10\\
EN031115\_031920&SB& 220.163&  52.25&  14.27&  29.67&   2.269&  0.8568&  0.3250& 117.19&   5.74& 157.37\\[-0.3ex]
&&&                            0.02&   0.02&   0.12&   0.030&  0.0025&  0.0014&   0.06&   0.03&   0.06\\
EN031115\_193751&SB& 220.844&  52.95&  14.53&  29.73&   2.276&  0.8577&  0.3239& 117.29&   5.65& 158.14\\[-0.3ex]
&&&                            0.07&   0.03&   0.03&   0.014&  0.0007&  0.0008&   0.14&   0.04&   0.14\\
EN031115\_195654&SB& 220.857&  53.26&  14.89&  29.91&   2.261&  0.8599&  0.3168& 118.16&   5.37& 159.03\\[-0.3ex]
&&&                            0.04&   0.08&   0.10&   0.029&  0.0022&  0.0011&   0.09&   0.10&   0.09\\
EN031115\_202247&N & 220.875&  51.46&  21.69&  29.92&   2.262&  0.8607&  0.3152& 298.38&   3.73& 159.24\\[-0.3ex]
&&&                            0.07&   0.06&   0.10&   0.030&  0.0021&  0.0012&   0.13&   0.08&   0.13\\
EN031115\_204226&SB& 220.888&  52.73&  14.89&  29.61&   2.273&  0.8564&  0.3264& 117.01&   5.10& 157.92\\[-0.3ex]
&&&                            0.03&   0.07&   0.04&   0.011&  0.0008&  0.0006&   0.07&   0.08&   0.07\\
EN031115\_212219&SB& 220.916&  52.99&  14.75&  29.75&   2.278&  0.8581&  0.3231& 117.37&   5.40& 158.30\\[-0.3ex]
&&&                            0.04&   0.03&   0.09&   0.027&  0.0020&  0.0010&   0.08&   0.04&   0.08\\
EN031115\_212455&SB& 220.918&  53.50&  14.24&  29.89&   2.254&  0.8588&  0.3182& 118.03&   6.26& 158.96\\[-0.3ex]
&&&                            0.02&   0.02&   0.08&   0.021&  0.0016&  0.0008&   0.04&   0.03&   0.04\\
EN031115\_213844&SB& 220.928&  52.78&  14.81&  29.60&   2.272&  0.8561&  0.3269& 116.95&   5.21& 157.90\\[-0.3ex]
&&&                            0.11&   0.03&   0.05&   0.023&  0.0013&  0.0014&   0.21&   0.06&   0.21\\
EN031115\_221917&SB& 220.956&  52.88&  14.43&  29.50&   2.251&  0.8539&  0.3289& 116.81&   5.66& 157.78\\[-0.3ex]
&&&                            0.02&   0.03&   0.09&   0.025&  0.0020&  0.0010&   0.05&   0.04&   0.05\\
EN031115\_221937&SB& 220.956&  52.99&  14.90&  29.71&   2.268&  0.8575&  0.3232& 117.39&   5.20& 158.37\\[-0.3ex]
&&&                            0.01&   0.03&   0.03&   0.009&  0.0007&  0.0004&   0.04&   0.04&   0.04\\
EN031115\_222446&SB& 220.960&  52.00&  14.45&  29.05&   2.273&  0.8487&  0.3438& 115.02&   5.19& 156.00\\[-0.3ex]
&&&                            0.04&   0.02&   0.08&   0.022&  0.0017&  0.0009&   0.07&   0.03&   0.07\\
EN031115\_225609&SB& 220.981&  53.44&  14.98&  30.02&   2.277&  0.8618&  0.3148& 118.32&   5.36& 159.32\\[-0.3ex]
&&&                            0.02&   0.07&   0.05&   0.015&  0.0011&  0.0007&   0.06&   0.09&   0.06\\
EN031115\_230149&SB& 220.985&  53.14&  14.53&  29.72&   2.269&  0.8574&  0.3235& 117.36&   5.70& 158.36\\[-0.3ex]
&&&                            0.03&   0.02&   0.09&   0.026&  0.0020&  0.0010&   0.06&   0.03&   0.06\\
EN031115\_232829&SB& 221.004&  53.09&  15.30&  29.83&   2.275&  0.8595&  0.3196& 117.78&   4.77& 158.80\\[-0.3ex]
&&&                            0.01&   0.02&   0.05&   0.013&  0.0010&  0.0005&   0.02&   0.02&   0.02\\
EN031115\_235911&S & 221.025&  51.85&  16.41&  28.35&   2.058&  0.8324&  0.3449& 115.81&   2.78& 156.87\\[-0.3ex]
&&&                            0.13&   0.18&   0.13&   0.032&  0.0030&  0.0022&   0.27&   0.21&   0.27\\
EN041115\_012728&SB& 221.087&  52.76&  14.08&  29.26&   2.245&  0.8503&  0.3360& 116.01&   5.95& 157.12\\[-0.3ex]
&&&                            0.02&   0.01&   0.03&   0.009&  0.0008&  0.0004&   0.04&   0.02&   0.04\\
EN041115\_020201&SB& 221.111&  52.55&  14.48&  29.24&   2.260&  0.8508&  0.3372& 115.83&   5.40& 156.95\\[-0.3ex]
&&&                            0.01&   0.02&   0.03&   0.008&  0.0006&  0.0004&   0.03&   0.02&   0.03\\
EN041115\_021111&SB& 221.117&  53.60&  14.77&  29.98&   2.274&  0.8610&  0.3160& 118.20&   5.65& 159.33\\[-0.3ex]
&&&                            0.05&   0.03&   0.04&   0.012&  0.0008&  0.0007&   0.09&   0.05&   0.09\\
EN041115\_021452&SB& 221.120&  53.16&  14.64&  29.67&   2.268&  0.8568&  0.3250& 117.19&   5.55& 158.33\\[-0.3ex]
&&&                            0.01&   0.01&   0.02&   0.005&  0.0003&  0.0002&   0.02&   0.02&   0.02\\
EN041115\_043317&SB& 221.216&  53.10&  14.11&  29.48&   2.269&  0.8539&  0.3315& 116.44&   6.10& 157.67\\[-0.3ex]
&&&                            0.02&   0.02&   0.04&   0.010&  0.0008&  0.0005&   0.04&   0.03&   0.04\\
EN041115\_044559&SB& 221.225&  53.31&  14.43&  29.66&   2.269&  0.8564&  0.3258& 117.10&   5.85& 158.34\\[-0.3ex]
&&&                            0.03&   0.02&   0.04&   0.010&  0.0008&  0.0005&   0.06&   0.02&   0.06\\
EN041115\_203853&SB& 221.888&  54.05&  14.80&  29.71&   2.260&  0.8569&  0.3234& 117.40&   5.64& 159.30\\[-0.3ex]
&&&                            0.02&   0.05&   0.05&   0.015&  0.0012&  0.0006&   0.06&   0.06&   0.06\\
EN041115\_210403&SB& 221.905&  53.92&  15.16&  29.64&   2.250&  0.8559&  0.3242& 117.34&   5.13& 159.26\\[-0.3ex]
&&&                            0.03&   0.11&   0.07&   0.021&  0.0016&  0.0009&   0.10&   0.13&   0.10\\
EN041115\_214032&SB& 221.931&  53.58&  14.86&  29.43&   2.267&  0.8535&  0.3321& 116.37&   5.30& 158.32\\[-0.3ex]
&&&                            0.03&   0.02&   0.07&   0.020&  0.0016&  0.0008&   0.06&   0.03&   0.06\\
EN041115\_215226&SB& 221.939&  53.82&  14.86&  29.44&   2.233&  0.8526&  0.3292& 116.83&   5.40& 158.79\\[-0.3ex]
&&&                            0.01&   0.01&   0.04&   0.009&  0.0008&  0.0004&   0.03&   0.02&   0.03\\
EN041115\_225243&SB& 221.981&  53.66&  14.61&  29.37&   2.258&  0.8523&  0.3335& 116.24&   5.61& 158.24\\[-0.3ex]
&&&                            0.02&   0.03&   0.10&   0.028&  0.0023&  0.0011&   0.05&   0.04&   0.05\\
EN041115\_231355&SB& 221.996&  53.95&  14.39&  29.51&   2.261&  0.8541&  0.3300& 116.64&   6.02& 158.64\\[-0.3ex]
&&&                            0.03&   0.05&   0.08&   0.022&  0.0018&  0.0009&   0.07&   0.06&   0.07\\
EN051115\_023102&SB& 222.133&  53.96&  14.96&  29.45&   2.241&  0.8530&  0.3295& 116.77&   5.32& 158.91\\[-0.3ex]
&&&                            0.01&   0.03&   0.02&   0.006&  0.0005&  0.0003&   0.03&   0.03&   0.03\\
EN051115\_183559&S & 222.804&  53.24&  13.90&  28.48&   2.259&  0.8398&  0.3618& 113.02&   5.93& 155.84\\[-0.3ex]
&&&                            0.01&   0.02&   0.01&   0.004&  0.0004&  0.0002&   0.02&   0.02&   0.02\\
EN051115\_185259&SB& 222.816&  54.30&  14.56&  29.26&   2.270&  0.8510&  0.3382& 115.65&   5.79& 158.49\\[-0.3ex]
&&&                            0.03&   0.03&   0.12&   0.035&  0.0028&  0.0012&   0.05&   0.03&   0.05\\
EN051115\_190203&SB& 222.823&  54.34&  14.66&  29.30&   2.270&  0.8516&  0.3371& 115.78&   5.70& 158.62\\[-0.3ex]
&&&                            0.02&   0.02&   0.03&   0.010&  0.0008&  0.0004&   0.03&   0.02&   0.03\\
EN051115\_203651&SB& 222.889&  54.18&  14.44&  29.11&   2.266&  0.8487&  0.3427& 115.15&   5.83& 158.06\\[-0.3ex]
&&&                            0.03&   0.03&   0.05&   0.013&  0.0010&  0.0006&   0.06&   0.03&   0.06\\
EN051115\_205304&N & 222.900&  53.87&  21.71&  29.26&   2.076&  0.8461&  0.3195& 298.63&   2.98& 161.50\\[-0.3ex]
&&&                            0.02&   0.01&   0.02&   0.005&  0.0004&  0.0003&   0.04&   0.01&   0.04\\
EN051115\_212802&SB& 222.924&  54.21&  14.94&  29.06&   2.235&  0.8475&  0.3409& 115.48&   5.26& 158.42\\[-0.3ex]
&&&                            0.02&   0.04&   0.04&   0.011&  0.0009&  0.0005&   0.05&   0.04&   0.05\\
EN051115\_213128&SB& 222.927&  54.20&  14.90&  29.14&   2.259&  0.8492&  0.3405& 115.43&   5.31& 158.38\\[-0.3ex]
&&&                            0.09&   0.16&   0.11&   0.033&  0.0024&  0.0017&   0.21&   0.19&   0.21\\
EN051115\_213433&SB& 222.929&  53.79&  14.60&  28.83&   2.255&  0.8448&  0.3499& 114.38&   5.43& 157.32\\[-0.3ex]
&&&                            0.01&   0.02&   0.03&   0.007&  0.0006&  0.0003&   0.02&   0.02&   0.02\\
EN051115\_220108&SB& 222.947&  53.92&  14.47&  28.88&   2.257&  0.8455&  0.3487& 114.50&   5.64& 157.47\\[-0.3ex]
&&&                            0.01&   0.01&   0.02&   0.007&  0.0006&  0.0003&   0.02&   0.02&   0.02\\
EN051115\_221253&SB& 222.956&  54.30&  14.77&  29.13&   2.252&  0.8488&  0.3404& 115.47&   5.50& 158.44\\[-0.3ex]
&&&                            0.05&   0.02&   0.03&   0.011&  0.0006&  0.0007&   0.11&   0.03&   0.11\\
EN051115\_221501&N & 222.957&  51.92&  21.71&  28.80&   2.253&  0.8454&  0.3484& 294.60&   3.32& 157.53\\[-0.3ex]
&&&                            0.01&   0.01&   0.08&   0.021&  0.0018&  0.0008&   0.02&   0.02&   0.02\\
EN051115\_221906&SB& 222.960&  54.28&  15.20&  29.16&   2.246&  0.8493&  0.3384& 115.72&   5.00& 158.70\\[-0.3ex]
&&&                            0.03&   0.02&   0.05&   0.014&  0.0012&  0.0006&   0.06&   0.03&   0.06\\
EN051115\_225625&SB& 222.986&  54.09&  14.54&  28.99&   2.262&  0.8471&  0.3458& 114.81&   5.65& 157.81\\[-0.3ex]
&&&                            0.01&   0.04&   0.03&   0.008&  0.0006&  0.0004&   0.04&   0.05&   0.04\\
EN051115\_225852&S & 222.988&  53.90&  15.19&  28.53&   2.150&  0.8373&  0.3498& 114.83&   4.74& 157.84\\[-0.3ex]
&&&                            0.01&   0.01&   0.04&   0.010&  0.0010&  0.0005&   0.02&   0.01&   0.02\\
EN051115\_231201&SB& 222.997&  54.25&  15.06&  29.15&   2.260&  0.8495&  0.3400& 115.48&   5.15& 158.49\\[-0.3ex]
&&&                            0.01&   0.00&   0.04&   0.010&  0.0009&  0.0004&   0.02&   0.01&   0.02\\
EN051115\_232719&N & 223.007&  55.29&  21.85&  29.42&   1.944&  0.8445&  0.3023& 301.24&   2.86& 164.22\\[-0.3ex]
&&&                            0.04&   0.03&   0.00&   0.005&  0.0001&  0.0005&   0.08&   0.05&   0.08\\
EN051115\_234939&SB& 223.023&  54.18&  14.47&  29.01&   2.263&  0.8473&  0.3454& 114.86&   5.76& 157.90\\[-0.3ex]
&&&                            0.01&   0.01&   0.05&   0.012&  0.0010&  0.0005&   0.02&   0.01&   0.02\\
EN051115\_235119&SB& 223.024&  54.53&  14.98&  29.35&   2.274&  0.8525&  0.3353& 115.96&   5.39& 159.00\\[-0.3ex]
&&&                            0.01&   0.01&   0.08&   0.023&  0.0018&  0.0009&   0.02&   0.02&   0.02\\
EN061115\_001740&SB& 223.042&  54.44&  14.66&  29.23&   2.275&  0.8508&  0.3394& 115.49&   5.70& 158.55\\[-0.3ex]
&&&                            0.01&   0.01&   0.03&   0.007&  0.0006&  0.0003&   0.01&   0.01&   0.01\\
EN061115\_002202&S & 223.045&  52.54&  15.27&  28.14&   2.261&  0.8362&  0.3704& 112.04&   4.10& 155.11\\[-0.3ex]
&&&                            0.02&   0.02&   0.09&   0.025&  0.0022&  0.0011&   0.05&   0.02&   0.05\\
EN061115\_003508&SB& 223.055&  54.24&  15.36&  29.15&   2.256&  0.8495&  0.3395& 115.55&   4.79& 158.62\\[-0.3ex]
&&&                            0.03&   0.05&   0.07&   0.018&  0.0014&  0.0008&   0.07&   0.06&   0.07\\
EN061115\_005009&S & 223.065&  54.31&  15.97&  29.26&   2.251&  0.8511&  0.3350& 116.08&   4.13& 159.16\\[-0.3ex]
&&&                            0.04&   0.13&   0.12&   0.033&  0.0027&  0.0015&   0.12&   0.15&   0.12\\
EN061115\_011233&SB& 223.081&  54.27&  14.94&  29.08&   2.256&  0.8484&  0.3420& 115.27&   5.27& 158.36\\[-0.3ex]
&&&                            0.01&   0.02&   0.04&   0.010&  0.0008&  0.0004&   0.03&   0.03&   0.03\\
EN061115\_011441&SB& 223.082&  53.57&  15.17&  28.72&   2.263&  0.8440&  0.3530& 113.99&   4.67& 157.09\\[-0.3ex]
&&&                            0.09&   0.05&   0.10&   0.029&  0.0022&  0.0015&   0.19&   0.06&   0.19\\
EN061115\_011623&SB& 223.083&  54.04&  14.54&  28.91&   2.266&  0.8462&  0.3483& 114.51&   5.59& 157.61\\[-0.3ex]
&&&                            0.02&   0.03&   0.12&   0.032&  0.0027&  0.0014&   0.06&   0.03&   0.06\\
EN061115\_025156&SB& 223.150&  54.42&  14.65&  29.17&   2.279&  0.8501&  0.3417& 115.22&   5.67& 158.38\\[-0.3ex]
&&&                            0.22&   0.03&   0.08&   0.041&  0.0021&  0.0027&   0.43&   0.09&   0.43\\
EN061115\_030548&S & 223.160&  54.69&  16.07&  29.54&   2.276&  0.8557&  0.3284& 116.74&   4.18& 159.92\\[-0.3ex]
&&&                            0.02&   0.03&   0.05&   0.014&  0.0011&  0.0007&   0.05&   0.04&   0.05\\
EN061115\_040629&SB& 223.202&  54.36&  14.47&  29.01&   2.263&  0.8473&  0.3456& 114.83&   5.80& 158.05\\[-0.3ex]
&&&                            0.01&   0.01&   0.01&   0.004&  0.0003&  0.0002&   0.03&   0.01&   0.03\\
EN061115\_164758&SB& 223.732&  54.85&  14.88&  28.94&   2.240&  0.8458&  0.3455& 114.93&   5.45& 158.68\\[-0.3ex]
&&&                            0.00&   0.01&   0.01&   0.002&  0.0002&  0.0001&   0.01&   0.01&   0.01\\
EN061115\_174311&SB& 223.771&  54.88&  15.22&  29.14&   2.279&  0.8499&  0.3421& 115.16&   5.11& 158.95\\[-0.3ex]
&&&                            0.01&   0.01&   0.04&   0.012&  0.0009&  0.0004&   0.03&   0.02&   0.03\\
EN071115\_015331&SB& 224.112&  54.80&  14.66&  28.71&   2.260&  0.8432&  0.3543& 113.84&   5.58& 157.97\\[-0.3ex]
&&&                            0.04&   0.02&   0.12&   0.030&  0.0027&  0.0014&   0.09&   0.03&   0.09\\
EN081115\_010613&SB& 225.083&  55.57&  14.79&  28.62&   2.271&  0.8422&  0.3584& 113.33&   5.58& 158.43\\[-0.3ex]
&&&                            0.04&   0.03&   0.12&   0.032&  0.0028&  0.0014&   0.08&   0.04&   0.08\\
EN081115\_033341&SB& 225.186&  55.71&  14.71&  28.57&   2.257&  0.8410&  0.3589& 113.33&   5.69& 158.53\\[-0.3ex]
&&&                            0.04&   0.06&   0.11&   0.028&  0.0025&  0.0014&   0.10&   0.07&   0.10\\
EN081115\_181258&SB& 225.799&  55.87&  14.75&  28.31&   2.264&  0.8377&  0.3674& 112.34&   5.58& 158.15\\[-0.3ex]
&&&                            0.01&   0.02&   0.01&   0.003&  0.0003&  0.0002&   0.02&   0.02&   0.02\\
EN081115\_202907&SB& 225.894&  55.47&  14.56&  28.02&   2.274&  0.8341&  0.3774& 111.17&   5.55& 157.08\\[-0.3ex]
&&&                            0.02&   0.02&   0.07&   0.021&  0.0018&  0.0008&   0.04&   0.02&   0.04\\
EN081115\_212839&SB& 225.935&  55.15&  14.83&  27.80&   2.260&  0.8307&  0.3827& 110.63&   5.12& 156.58\\[-0.3ex]
&&&                            0.01&   0.01&   0.02&   0.007&  0.0006&  0.0003&   0.01&   0.01&   0.01\\
EN081115\_234417&SB& 226.030&  55.75&  14.48&  28.02&   2.254&  0.8331&  0.3761& 111.39&   5.72& 157.43\\[-0.3ex]
&&&                            0.03&   0.02&   0.08&   0.021&  0.0019&  0.0009&   0.05&   0.02&   0.05\\
EN091115\_001801&SB& 226.053&  55.15&  15.25&  27.82&   2.272&  0.8318&  0.3822& 110.63&   4.67& 156.71\\[-0.3ex]
&&&                            0.02&   0.03&   0.16&   0.041&  0.0038&  0.0017&   0.06&   0.03&   0.06\\
EN091115\_003545&SB& 226.066&  55.43&  15.02&  27.91&   2.261&  0.8323&  0.3791& 111.02&   5.01& 157.11\\[-0.3ex]
&&&                            0.05&   0.02&   0.03&   0.011&  0.0008&  0.0007&   0.09&   0.03&   0.09\\
EN091115\_011246&SB& 226.092&  55.46&  14.33&  27.72&   2.237&  0.8282&  0.3842& 110.55&   5.71& 156.66\\[-0.3ex]
&&&                            0.04&   0.10&   0.11&   0.028&  0.0026&  0.0014&   0.11&   0.10&   0.11\\
EN091115\_011650&SB& 226.094&  55.36&  15.19&  27.87&   2.258&  0.8318&  0.3799& 110.94&   4.80& 157.06\\[-0.3ex]
&&&                            0.02&   0.03&   0.12&   0.029&  0.0028&  0.0013&   0.04&   0.04&   0.04\\
EN091115\_032502&SB& 226.184&  55.99&  15.27&  28.27&   2.277&  0.8379&  0.3691& 112.08&   5.00& 158.28\\[-0.3ex]
&&&                            0.02&   0.02&   0.02&   0.007&  0.0006&  0.0004&   0.04&   0.03&   0.04\\
EN091115\_041944&N & 226.222&  56.51&  21.89&  29.90&   2.387&  0.8639&  0.3248& 296.79&   2.51& 162.97\\[-0.3ex]
&&&                            0.03&   0.02&   0.06&   0.017&  0.0012&  0.0007&   0.05&   0.03&   0.05\\
EN101115\_212402&SB& 227.942&  56.75&  15.16&  27.47&   2.242&  0.8252&  0.3918& 109.66&   5.03& 157.62\\[-0.3ex]
&&&                            0.03&   0.03&   0.05&   0.015&  0.0014&  0.0007&   0.07&   0.03&   0.07\\
EN101115\_235401&SB& 228.047&  57.00&  15.45&  27.66&   2.260&  0.8287&  0.3870& 110.11&   4.84& 158.18\\[-0.3ex]
&&&                            0.04&   0.02&   0.03&   0.009&  0.0007&  0.0006&   0.08&   0.03&   0.08\\
EN111115\_004713&S & 228.084&  56.33&  13.94&  27.01&   2.256&  0.8187&  0.4090& 107.68&   5.99& 155.77\\[-0.3ex]
&&&                            0.01&   0.02&   0.08&   0.020&  0.0020&  0.0009&   0.03&   0.02&   0.03\\
EN111115\_031037&SB& 228.184&  57.37&  14.70&  27.64&   2.248&  0.8275&  0.3880& 110.06&   5.71& 158.26\\[-0.3ex]
&&&                            0.12&   0.23&   0.06&   0.026&  0.0017&  0.0019&   0.29&   0.24&   0.29\\
EN111115\_181413&SB& 228.815&  57.27&  14.96&  27.30&   2.266&  0.8236&  0.3997& 108.66&   5.27& 157.50\\[-0.3ex]
&&&                            0.06&   0.07&   0.08&   0.025&  0.0021&  0.0011&   0.13&   0.07&   0.13\\
EN111115\_181509&SB& 228.815&  57.11&  15.06&  27.20&   2.262&  0.8223&  0.4021& 108.41&   5.10& 157.24\\[-0.3ex]
&&&                            0.03&   0.03&   0.00&   0.004&  0.0002&  0.0003&   0.05&   0.03&   0.05\\
EN111115\_184540&N & 228.837&  58.35&  22.60&  29.04&   2.286&  0.8494&  0.3443& 294.88&   2.78& 163.68\\[-0.3ex]
&&&                            0.13&   0.02&   0.06&   0.026&  0.0015&  0.0016&   0.25&   0.03&   0.25\\
EN111115\_203917&SB& 228.916&  56.93&  14.97&  27.08&   2.277&  0.8212&  0.4072& 107.77&   5.09& 156.71\\[-0.3ex]
&&&                            0.12&   0.04&   0.05&   0.023&  0.0015&  0.0015&   0.24&   0.05&   0.24\\
EN111115\_233243&SB& 229.037&  57.63&  15.13&  27.34&   2.250&  0.8235&  0.3970& 109.02&   5.19& 158.08\\[-0.3ex]
&&&                            0.03&   0.01&   0.04&   0.012&  0.0011&  0.0006&   0.06&   0.02&   0.06\\
EN121115\_004717&SB& 229.089&  57.10&  15.23&  26.97&   2.242&  0.8183&  0.4074& 107.90&   4.85& 157.01\\[-0.3ex]
&&&                            0.03&   0.04&   0.07&   0.017&  0.0017&  0.0008&   0.07&   0.04&   0.07\\
EN121115\_232341&S & 230.037&  56.31&  14.51&  25.93&   2.262&  0.8041&  0.4431& 103.82&   5.00& 153.88\\[-0.3ex]
&&&                            0.02&   0.05&   0.05&   0.014&  0.0014&  0.0007&   0.05&   0.04&   0.05\\
EN131115\_002058&SB& 230.077&  58.58&  15.58&  27.39&   2.269&  0.8251&  0.3968& 108.96&   4.94& 159.05\\[-0.3ex]
&&&                            0.02&   0.06&   0.09&   0.023&  0.0022&  0.0011&   0.06&   0.06&   0.06\\
EN131115\_004858&SB& 230.097&  58.41&  15.35&  27.27&   2.275&  0.8236&  0.4013& 108.42&   5.09& 158.54\\[-0.3ex]
&&&                            0.02&   0.02&   0.09&   0.023&  0.0022&  0.0011&   0.05&   0.02&   0.05\\
EN131115\_015008&SB& 230.139&  58.33&  14.98&  27.15&   2.277&  0.8219&  0.4057& 107.93&   5.41& 158.09\\[-0.3ex]
&&&                            0.01&   0.01&   0.03&   0.009&  0.0008&  0.0004&   0.03&   0.01&   0.03\\
EN131115\_042559&SB& 230.248&  58.11&  15.37&  26.98&   2.270&  0.8195&  0.4097& 107.51&   4.91& 157.77\\[-0.3ex]
&&&                            0.09&   0.02&   0.08&   0.024&  0.0021&  0.0015&   0.17&   0.03&   0.17\\
EN161115\_193458&S & 233.906&  60.56&  14.63&  25.65&   2.162&  0.7942&  0.4450& 104.03&   5.66& 157.95\\[-0.3ex]
&&&                            0.03&   0.01&   0.02&   0.007&  0.0007&  0.0004&   0.06&   0.01&   0.06\\
EN161115\_213048&SB& 233.987&  60.23&  15.09&  25.86&   2.257&  0.8025&  0.4457& 103.50&   5.20& 157.51\\[-0.3ex]
&&&                            0.01&   0.01&   0.04&   0.009&  0.0010&  0.0004&   0.02&   0.01&   0.02\\
EN161115\_222246&S & 234.023&  67.52&  16.82&  29.10&   1.957&  0.8372&  0.3185& 119.17&   6.26& 173.21\\[-0.3ex]
&&&                            0.06&   0.02&   0.09&   0.018&  0.0019&  0.0011&   0.12&   0.03&   0.12\\
EN171115\_020907&SB& 234.182&  60.80&  15.58&  26.20&   2.273&  0.8083&  0.4357& 104.54&   4.94& 158.74\\[-0.3ex]
&&&                            0.02&   0.08&   0.00&   0.004&  0.0001&  0.0004&   0.06&   0.08&   0.06\\
EN171115\_022102&SB& 234.190&  60.40&  15.51&  25.91&   2.260&  0.8036&  0.4440& 103.68&   4.84& 157.89\\[-0.3ex]
&&&                            0.02&   0.02&   0.04&   0.010&  0.0010&  0.0005&   0.03&   0.02&   0.03\\
EN231115\_012005&N & 240.202&  66.46&  24.65&  26.27&   2.147&  0.8035&  0.4219& 286.63&   2.93& 166.80\\[-0.3ex]
&&&                            0.03&   0.01&   0.03&   0.009&  0.0009&  0.0005&   0.06&   0.01&   0.06\\
EN231115\_224311&S & 241.102&  69.19&  16.03&  25.73&   1.992&  0.7853&  0.4277& 106.76&   5.83& 167.88\\[-0.3ex]
&&&                            0.16&   0.02&   0.06&   0.022&  0.0019&  0.0020&   0.31&   0.04&   0.31\\
EN281115\_195251&S & 246.038&  70.87&  16.55&  25.18&   2.300&  0.7936&  0.4747&  99.91&   5.12& 165.97\\[-0.3ex]
&&&                            0.03&   0.02&   0.05&   0.016&  0.0016&  0.0006&   0.05&   0.02&   0.05\\
\end{longtable}
\footnotetext[1]{N -- Northern (13), S -- Southern (18), SB -- Southern new branch  (113)}

\end{longtab}

\begin{longtab}
\begin{longtable}{lcrrrrrrrll}
\caption{Physical data of 2015 Taurid fireballs. The entry velocity, heights of beginning, maximum brightness and end, average zenith distance
of the radiant, photometric mass, maximum absolute magnitude, PE coefficient, and classification according to PE are given. Code of each fireball contains also date (in ddmmyy format) and GMT time corresponding to beginning rounded to whole second (in hhmmss format)}
\label{tbphysical}\\
\hline\hline \\[-3mm]
Code & \multicolumn {1}{c}{Branch} & $v_\infty$ & $H_{\rm beg}$ & $H_{\rm max}$& $H_{\rm end}$ & $Z_{\rm rad}$ & Mass & Mag & PE & Type \\
&& km/s & km & km & km & deg & kg & \\
\hline \\[-2mm]
\endfirsthead
\caption{continued.}\\
\hline\hline \\[-3mm]
Code & \multicolumn {1}{c}{Branch} & $v_\infty$ & $H_{\rm beg}$ & $H_{\rm max}$& $H_{\rm end}$ & $Z_{\rm rad}$ & Mass & Mag & PE & Type  \\
&& km/s & km & km & km & deg & kg & \\
\hline \\[-2mm]
\endhead
\hline
\endfoot
EN231015\_204348&S &36.98& 106.3&  73.2&  59.7&  54.7&0.0055&  -7.5& -4.52& I   \\
EN231015\_211327&N &32.30&  96.9&  64.9&  57.8&  38.2&0.0021&  -4.2& -4.51& I   \\
EN241015\_004546&N &34.25& 103.9&  65.7&  64.8&  31.4&0.0022&  -8.2& -4.89& II  \\
EN241015\_185031&S &30.58&  98.8&  73.9&  65.6&  67.3& 0.012&  -6.3& -4.87& II  \\
EN251015\_022301&SB&34.30& 104.5&  77.6&  69.7&  44.6&  0.18& -12.0& -5.87& IIIB\\
EN251015\_031725&SB&33.40& 108.3&  71.1&  66.9&  52.3&0.0011&  -4.9& -4.71& II  \\
EN261015\_213736&SB&34.38&  91.9&  72.2&  64.0&  43.4&0.0048&  -6.4& -4.90& II  \\
EN261015\_224031&SB&33.95&  94.6&  72.1&  61.3&  38.2&0.0041&  -6.5& -4.78& II  \\
EN271015\_220749&SB&33.75&  94.3&  71.6&  62.4&  41.9&  0.20& -10.5& -5.52& IIIA\\
EN281015\_011855&S &28.87&  90.1&  77.9&  65.8&  42.8&0.0023&  -5.0& -4.98& II  \\
EN301015\_222401&SB&32.98& 104.0&  67.5&  65.5&  38.9&0.0047&  -8.9& -5.04& II  \\
EN311015\_002325&N &30.38& 102.7&  68.1&  59.3&  32.0&0.0037&  -6.3& -4.77& II  \\
EN311015\_023900&SB&32.63& 105.5&  80.9&  68.5&  47.5&0.0055&  -6.3& -5.17& II  \\
EN311015\_025717&SB&32.63&  99.8&  71.7&  57.9&  53.3&  0.21& -11.5& -5.19& II  \\
EN311015\_172431&SB&33.05& 102.0&  87.4&  81.9&  76.5&0.0010&  -3.5& -5.12& II  \\
EN311015\_180520&SB&33.06& 114.7&  80.8&  57.6&  72.1&  1300& -18.6& -6.31& IIIB\\
EN311015\_182902&SB&32.61&  99.0&  79.0&  74.0&  69.8&0.0005&  -4.0& -4.66& II  \\
EN311015\_185530&SB&32.80& 103.7&  77.1&  69.4&  63.7&0.0006&  -3.0& -4.55& II  \\
EN311015\_192126&S &31.90&  99.9&  69.9&  69.5&  62.6&0.0006&  -3.5& -4.62& II  \\
EN311015\_200534&SB&32.76& 100.3&  76.7&  67.2&  52.2&0.0020&  -5.1& -4.85& II  \\
EN311015\_202117&SB&32.43& 105.7&  60.3&  55.2&  53.5&0.0070&  -8.0& -4.43& I   \\
EN311015\_211904&SB&33.08& 102.8&  75.5&  71.3&  46.8& 0.039& -10.2& -5.68& IIIB\\
EN311015\_230919&SB&32.78& 100.2&  69.9&  64.1&  33.6&0.0006&  -3.8& -4.61& II  \\
EN311015\_231301&SB&32.56& 120.0&  74.4&  57.3&  36.9&    34& -15.8& -6.19& IIIB\\
EN011115\_013625&SB&32.41&  98.8&  69.0&  58.2&  41.0& 0.021&  -9.5& -4.92& II  \\
EN011115\_033911&N &32.92& 101.2&  74.7&  71.6&  54.0&0.0081&  -9.4& -5.33& IIIA\\
EN011115\_174410&SB&32.64& 102.3&  71.7&  67.2&  75.6&0.0089&  -6.6& -4.62& II  \\
EN011115\_183646&S &32.55&  95.0&  82.3&  80.1&  65.8&0.0002&  -2.4& -5.00& II  \\
EN011115\_191104&SB&32.24& 104.3&  78.4&  73.8&  63.4&0.0049&  -4.1& -5.00& II  \\
EN011115\_200918&SB&32.47&  93.2&  64.5&  61.3&  53.2&0.0030&  -6.9& -4.60& II  \\
EN011115\_223909&SB&32.52& 102.2&  76.5&  62.8&  37.3&0.0021&  -4.6& -4.72& II  \\
EN011115\_234207&SB&32.60&  96.3&  71.5&  57.6&  31.3&0.0083&  -7.5& -4.77& II  \\
EN021115\_020950&SB&32.05&  99.8&  77.0&  69.7&  45.1&0.0020&  -6.1& -5.09& II  \\
EN021115\_021740&SB&32.00&  94.0&  74.9&  68.2&  45.1&0.0008&  -4.2& -4.83& II  \\
EN021115\_022525&SB&32.38& 107.5&  75.2&  63.3&  45.3&  0.20& -10.9& -5.57& IIIA\\
EN021115\_024553&N &29.72&  99.9&  59.3&  53.9&  47.5& 0.027&  -7.8& -4.68& II  \\
EN021115\_182450&SB&31.93&  94.9&  72.7&  62.7&  71.2& 0.031&  -7.8& -4.76& II  \\
EN021115\_195540&SB&32.46& 100.8&  79.0&  75.1&  55.8&0.0018&  -5.3& -5.25& IIIA\\
EN021115\_201534&SB&32.19& 102.9&  76.5&  72.1&  53.6&0.0002&  -2.1& -4.73& II  \\
EN021115\_205431&N &33.00& 111.9&  76.5&  73.8&  42.9&0.0003&  -3.2& -4.98& II  \\
EN021115\_213614&SB&32.17&  97.5&  70.8&  60.7&  45.4&  0.15& -11.3& -5.37& IIIA\\
EN021115\_215818&SB&32.16&  99.5&  77.0&  73.3&  40.8&0.0017&  -6.7& -5.31& IIIA\\
EN021115\_220435&SB&32.11& 108.6&  70.8&  63.6&  39.3&0.0010&  -6.1& -4.68& II  \\
EN021115\_232112&SB&31.80&  94.4&  54.1&  48.4&  35.5& 0.015&  -9.3& -4.39& I   \\
EN021115\_234348&S &30.66& 100.5&  78.5&  70.9&  32.8&0.0001&  -1.9& -4.78& II  \\
EN021115\_235259&SB&31.98& 100.5&  62.9&  58.9&  33.2&0.0031&  -7.5& -4.68& II  \\
EN031115\_002007&SB&31.78& 101.9&  73.3&  58.7&  35.1&0.0004&  -3.0& -4.28& I   \\
EN031115\_011247&SB&32.10&  98.8&  81.0&  67.4&  37.9&0.0003&  -2.6& -4.67& II  \\
EN031115\_012404&SB&31.87&  97.6&  71.6&  61.1&  41.9&0.0015&  -4.3& -4.59& II  \\
EN031115\_025102&SB&31.85&  99.8&  73.0&  69.1&  48.4&0.0004&  -3.0& -4.71& II  \\
EN031115\_031920&SB&31.43&  97.5&  81.5&  71.3&  56.1&0.0005&  -3.2& -4.80& II  \\
EN031115\_193751&SB&31.99& 102.6&  80.3&  69.9&  57.8&0.0053&  -6.4& -5.12& II  \\
EN031115\_195654&SB&32.13&  97.7&  69.1&  68.4&  54.6&0.0055&  -9.1& -5.09& II  \\
EN031115\_202247&N &32.12&  94.4&  79.1&  72.8&  44.5&0.0008&  -4.2& -5.12& II  \\
EN031115\_204226&SB&31.84& 100.9&  65.9&  59.2&  51.0&0.0013&  -4.9& -4.37& I   \\
EN031115\_212219&SB&31.92&  96.5&  69.6&  68.1&  42.1&0.0017&  -6.7& -4.99& II  \\
EN031115\_212455&SB&32.04&  99.1&  64.8&  59.8&  41.7& 0.013&  -9.8& -4.92& II  \\
EN031115\_213844&SB&31.77& 101.5&  72.7&  66.8&  42.9&0.0029&  -6.5& -5.01& II  \\
EN031115\_221917&SB&31.63& 101.6&  83.8&  74.6&  38.5&0.0003&  -2.9& -5.08& II  \\
EN031115\_221937&SB&31.82&  98.9&  64.6&  59.9&  39.6&0.0047&  -7.7& -4.77& II  \\
EN031115\_222446&SB&31.19& 100.8&  72.9&  69.4&  36.2&0.0003&  -2.6& -4.78& II  \\
EN031115\_225609&SB&32.06&  95.9&  72.0&  61.7&  35.7&0.0008&  -3.8& -4.57& I   \\
EN031115\_230149&SB&31.78&  99.1&  74.5&  66.1&  33.1&0.0010&  -4.7& -4.85& II  \\
EN031115\_232829&SB&31.85&  98.6&  69.5&  59.2&  34.9&0.0030&  -7.0& -4.68& II  \\
EN031115\_235911&S &30.41&  98.1&  76.1&  66.8&  30.1& 0.012&  -8.4& -5.40& IIIA\\
EN041115\_012728&SB&31.16& 106.5&  63.7&  48.1&  39.5&0.0018&  -4.1& -3.97& I   \\
EN041115\_020201&SB&31.10&  98.1&  63.9&  58.1&  45.9&0.0033&  -6.3& -4.56& I   \\
EN041115\_021111&SB&31.80& 100.2&  71.7&  67.2&  42.5&  0.44& -12.7& -5.96& IIIB\\
EN041115\_021452&SB&31.50& 102.8&  68.4&  63.8&  45.1&0.0076&  -8.8& -5.01& II  \\
EN041115\_043317&SB&31.22& 101.8&  83.0&  75.3&  65.7&0.0005&  -3.0& -4.88& II  \\
EN041115\_044559&SB&31.39& 100.1&  82.7&  75.4&  65.8&0.0005&  -2.6& -4.89& II  \\
EN041115\_203853&SB&31.90&  95.0&  74.9&  60.2&  48.2&  0.21& -10.6& -5.39& IIIA\\
EN041115\_210403&SB&31.81&  98.7&  64.7&  62.9&  45.1& 0.031& -10.1& -5.21& II  \\
EN041115\_214032&SB&31.58&  94.1&  68.2&  56.2&  38.3& 0.016&  -9.8& -4.81& II  \\
EN041115\_215226&SB&31.60&  93.9&  64.2&  52.6&  40.1&0.0032&  -4.2& -4.29& I   \\
EN041115\_225243&SB&31.45&  97.3&  67.9&  60.4&  35.4&0.0015&  -5.6& -4.61& II  \\
EN041115\_231355&SB&31.56&  92.3&  71.7&  63.5&  35.5& 0.019&  -9.2& -5.25& IIIA\\
EN051115\_023102&SB&31.27&  98.0&  65.7&  56.9&  47.8& 0.011&  -8.8& -4.69& II  \\
EN051115\_183559&S &30.84&  99.9&  68.3&  65.3&  67.2&0.0087&  -7.8& -4.80& II  \\
EN051115\_185259&SB&31.57&  99.7&  79.6&  75.7&  64.4&0.0004&  -3.3& -4.88& II  \\
EN051115\_190203&SB&31.61&  99.6&  78.5&  72.5&  63.0&0.0008&  -3.6& -4.85& II  \\
EN051115\_203651&SB&31.35& 101.1&  71.7&  66.3&  47.6&0.0010&  -6.3& -4.76& II  \\
EN051115\_205304&N &31.46& 103.5&  70.0&  57.4&  38.5& 0.093& -10.4& -5.20& II  \\
EN051115\_212802&SB&31.27&  96.5&  72.3&  65.5&  43.7&0.0016&  -4.7& -4.84& II  \\
EN051115\_213128&SB&31.32&  91.4&  70.8&  66.3&  43.7& 0.090& -10.5& -5.62& IIIA\\
EN051115\_213433&SB&31.03& 105.9&  62.2&  58.1&  41.0&0.0038&  -7.4& -4.63& II  \\
EN051115\_220108&SB&31.05&  99.3&  70.8&  66.0&  38.9& 0.050& -10.0& -5.54& IIIA\\
EN051115\_221253&SB&31.28& 102.4&  64.7&  60.9&  36.6& 0.010&  -9.6& -4.99& II  \\
EN051115\_221501&N &30.95& 107.8&  77.0&  70.8&  29.9&0.0005&  -3.9& -5.02& II  \\
EN051115\_221906&SB&31.28&  96.7&  76.3&  67.6&  35.5&0.0006&  -4.0& -4.84& II  \\
EN051115\_225625&SB&31.10&  99.9&  76.7&  71.2&  35.2&0.0003&  -3.1& -4.90& II  \\
EN051115\_225852&S &30.66&  98.7&  73.3&  63.7&  34.9&0.0019&  -5.7& -4.85& II  \\
EN051115\_231201&SB&31.22& 105.3&  72.0&  62.1&  32.3&  0.11& -11.0& -5.52& IIIA\\
EN051115\_232719&N &31.48& 101.0&  71.0&  64.6&  25.4&0.0018&  -6.2& -4.94& II  \\
EN051115\_234939&SB&31.05& 102.5&  61.5&  57.0&  34.2&0.0026&  -6.5& -4.56& I   \\
EN051115\_235119&SB&31.37&  99.3&  73.7&  63.7&  33.4&0.0019&  -6.8& -4.85& II  \\
EN061115\_001740&SB&31.20& 104.8&  68.5&  61.1&  35.2&0.0044&  -7.8& -4.86& II  \\
EN061115\_002202&S &30.20& 102.2&  62.9&  54.0&  33.4&0.0009&  -4.3& -4.22& I   \\
EN061115\_003508&SB&31.10& 100.4&  66.6&  65.3&  36.5&0.0052&  -9.1& -5.11& II  \\
EN061115\_005009&S &31.20&  94.2&  74.4&  64.2&  37.9&0.0045&  -7.8& -5.01& II  \\
EN061115\_011233&SB&31.02&  98.9&  60.1&  57.5&  38.5&0.0031&  -8.1& -4.58& I   \\
EN061115\_011441&SB&30.67&  98.9&  83.6&  71.9&  37.0&0.0003&  -3.4& -4.94& II  \\
EN061115\_011623&SB&30.85&  97.2&  69.0&  64.5&  39.9&0.0004&  -2.9& -4.57& I   \\
EN061115\_025156&SB&30.98&  94.5&  77.5&  73.2&  50.1&0.0026&  -5.0& -5.31& IIIA\\
EN061115\_030548&S &31.33&  98.8&  75.9&  71.7&  51.5&0.0020&  -7.3& -5.15& II  \\
EN061115\_040629&SB&30.78& 101.1&  71.1&  65.0&  63.3&0.0100&  -8.2& -4.89& II  \\
EN061115\_164758&SB&31.28& 104.0&  79.7&  74.0&  83.3&0.0098&  -5.1& -4.64& II  \\
EN061115\_174311&SB&31.47&  99.6&  79.8&  76.5&  73.3&0.0018&  -4.2& -4.99& II  \\
EN071115\_015331&SB&30.60& 104.4&  76.7&  59.0&  43.7&   3.6& -13.8& -5.92& IIIB\\
EN081115\_010613&SB&30.59&  92.4&  78.7&  71.5&  37.9&0.0014&  -5.9& -5.23& II  \\
EN081115\_033341&SB&30.41&  96.4&  72.3&  70.0&  57.5&0.0017&  -5.3& -4.95& II  \\
EN081115\_181258&SB&30.69&  99.1&  65.9&  65.6&  70.2& 0.049&  -9.8& -5.05& II  \\
EN081115\_202907&SB&30.35& 100.2&  68.1&  66.6&  50.0&0.0020&  -6.3& -4.90& II  \\
EN081115\_212839&SB&30.08& 103.1&  68.3&  57.6&  41.3&0.0022&  -5.0& -4.52& I   \\
EN081115\_234417&SB&30.10&  99.4&  70.9&  63.4&  33.0&0.0062&  -7.9& -5.08& II  \\
EN091115\_001801&SB&29.90& 102.9&  77.6&  72.4&  33.3&0.0009&  -5.2& -5.25& II  \\
EN091115\_003545&SB&29.97&  96.2&  69.0&  53.2&  33.5& 0.041&  -9.3& -4.85& II  \\
EN091115\_011246&SB&29.73&  88.7&  75.7&  70.3&  40.5&0.0015&  -5.0& -5.15& II  \\
EN091115\_011650&SB&29.88&  98.4&  76.5&  71.4&  38.5&0.0003&  -3.3& -4.96& II  \\
EN091115\_032502&SB&30.13&  99.6&  78.8&  73.4&  54.7&0.0015&  -4.5& -5.18& IIIA\\
EN091115\_041944&N &31.62&  99.3&  73.9&  67.8&  61.1&0.0052&  -8.5& -4.95& II  \\
EN101115\_212402&SB&29.77& 102.4&  63.5&  58.4&  41.5& 0.011&  -9.6& -4.86& II  \\
EN101115\_235401&SB&29.75&  98.1&  71.7&  66.3&  33.5&  0.12& -10.9& -5.78& IIIB\\
EN111115\_004713&S &29.11&  97.3&  63.9&  56.9&  35.8& 0.022&  -9.6& -4.97& II  \\
EN111115\_031037&SB&29.55&  95.0&  71.6&  65.9&  54.8&  0.91& -13.5& -5.92& IIIB\\
EN111115\_181413&SB&29.76&  97.2&  72.0&  62.1&  66.6&  0.23& -11.4& -5.26& IIIA\\
EN111115\_181509&SB&29.66&  98.5&  74.1&  69.2&  65.3&  0.27& -11.8& -5.70& IIIB\\
EN111115\_184540&N &31.36& 102.0&  65.9&  72.0&  59.4&0.0010&  -3.3& -4.92& II  \\
EN111115\_203917&SB&29.47&  93.5&  77.6&  74.5&  46.5&0.0010&  -3.8& -5.29& IIIA\\
EN111115\_233243&SB&29.51&  98.9&  68.1&  60.1&  32.4& 0.014&  -8.3& -5.07& II  \\
EN121115\_004717&SB&29.07&  96.9&  74.4&  71.6&  36.9&0.0008&  -5.4& -5.17& II  \\
EN121115\_232341&S &28.20&  95.7&  74.2&  66.6&  33.8&0.0005&  -3.4& -4.83& II  \\
EN131115\_002058&SB&29.48&  97.4&  77.4&  66.1&  36.0&0.0074&  -8.2& -5.25& IIIA\\
EN131115\_004858&SB&29.34&  94.5&  79.2&  67.7&  35.8&0.0003&  -2.3& -4.76& II  \\
EN131115\_015008&SB&29.17& 103.1&  74.7&  61.2&  43.8&0.0014&  -5.0& -4.63& II  \\
EN131115\_042559&SB&28.90&  93.7&  75.9&  74.2&  64.1&0.0009&  -3.5& -5.01& II  \\
EN161115\_193458&S &28.19& 101.5&  60.8&  56.0&  53.5&0.0079&  -6.1& -4.58& I   \\
EN161115\_213048&SB&28.28& 100.4&  70.5&  59.4&  39.6& 0.017& -10.1& -5.05& II  \\
EN161115\_222246&S &31.25&  97.9&  74.3&  69.1&  35.2&0.0004&  -3.6& -4.86& II  \\
EN171115\_020907&SB&28.26&  93.5&  68.8&  62.0&  46.6&0.0062&  -6.2& -4.93& II  \\
EN171115\_022102&SB&27.98&  95.1&  66.2&  59.0&  47.5& 0.049& -10.6& -5.16& II  \\
EN231115\_012005&N &28.37&  93.2&  64.7&  57.5&  33.2&0.0040&  -5.8& -4.72& II  \\
EN231115\_224311&S &28.06&  96.8&  68.5&  57.0&  31.7& 0.074& -10.3& -5.25& II  \\
EN281115\_195251&S &27.74& 108.0&  78.2&  64.7&  50.3&0.0016&  -4.1& -4.80& II  \\
\end{longtable}
\end{longtab}

\end{document}